\documentclass[10pt]{article}
\pdfoutput=1
\usepackage{jheppub}

\usepackage{slashed}
\usepackage{color, verbatim}
\usepackage{latexsym}
\usepackage{epsfig}
\usepackage{amsmath,accents}
\usepackage{mathrsfs}

\usepackage{amssymb}
\usepackage{graphicx}
\usepackage{bm}
\usepackage{stackrel}

\usepackage[font={small}]{caption}

\usepackage{hyperref}
\usepackage{epstopdf}
\usepackage{cancel}

\definecolor{myred}{rgb}{0.7, 0, 0}
\definecolor{myblue}{rgb}{0, 0, 0.7}
\definecolor{mygreen}{rgb}{0.04, 0.7, 0.5}
\hypersetup{colorlinks,citecolor=myred,linkcolor=myblue,urlcolor=myblue,linktocpage=true}

\makeatletter

\@addtoreset{equation}{section}
\makeatother

\def \bse {\begin{subequations} \begin{eqnarray}}
\def \ese {\end{eqnarray} \end{subequations}}
\def \be {\begin{equation}}
\def \ee {\end{equation}}
\def \bi {\begin{enumerate}}
\def \ei {\end{enumerate}}
\def \ba {\begin{aligned}}
\def \ea {\end{aligned}}
\def \bse {\begin{subequations} \begin{eqnarray}}
\def \ese {\end{eqnarray} \end{subequations}}

\def \Mp {M_\text{pl}}

\begin{document}

\thispagestyle{empty}

\begin{center}

\vspace{.5cm}

{\Large\sc
Numerical study of the Schwinger effect \\ 
in axion inflation
}\\

\vspace{1.cm}

\textsc{
Yann Cado, 
Mariano Quir\'os
}\\

\vspace{.5cm}

{\em {Institut de F\'{\i}sica d'Altes Energies (IFAE) and\\ The Barcelona Institute of  Science and Technology (BIST),\\ Campus UAB, 08193 Bellaterra, Barcelona, Spain
}}

\end{center}

\vspace{0.8cm}

\centerline{\bf Abstract}
\vspace{2 mm}

\begin{quote}\small
Previous studies demonstrate that the inflaton, when coupled to the hypercharge Chern-Simons density, can source an explosive production of helical hypermagnetic fields.
Then, in the absence of fermion production, those fields have the capability of preheating the Universe after inflation and triggering a successful baryogenesis mechanism at the electroweak phase transition.
In the presence of fermion production however, we expect a strong damping of the gauge fields production from the fermion backreaction, a phenomenon called Schwinger effect, thus jeopardizing their original capabilities. Using numerical methods we study the backreaction on the generated gauge fields and revisit the processes of gauge preheating and baryogenesis in the presence of the Schwinger effect. We have found that gauge preheating is very unlikely, while still having a sizable window in the parameter space to achieve the baryon asymmetry of the Universe at the electroweak phase transition.
\end{quote}
 
\vfill

\newpage

\tableofcontents

\newpage

\section{Introduction}
\label{sec:introduction}
Cosmological inflation~\cite{Guth:1980zm,Linde:1981mu,Albrecht:1982wi} is nowadays a well-established paradigm to solve the classical (flatness, horizon,~\dots) problems of the Standard Cosmological Model, and to generate the primordial density perturbations giving rise to the present Universe structure. The achievements of cosmological inflation usually require the presence of one (or several) scalar field --the inflaton-- giving rise to physics beyond the Standard Model (SM) of Particle Physics (BSM). In this way, along with the classical problems of the SM (hierarchy problem, baryogenesis, strong CP-problem, dark matter,...), cosmological inflation provides yet another motivation for BSM physics. 

Although the existence of a period of cosmological inflation is pretty well stablished by observational cosmological data~\cite{Planck:2018jri}, there is no consensus on a detailed model. An interesting candidate for the inflaton is a pseudoscalar field $\phi$, denoted in this paper as \textit{axionlike particle}\footnote{With an (obvious) abuse of language, we are identifying in this paper axions with axionlike (or pseudoscalar) particles, which allow a wider choice for inflationary potentials~\cite{Adshead:2019lbr,Adshead:2019igv}.}, 
%
%
%
which can then couple to the Chern-Simons density $F_{\mu\nu}\tilde F^{\mu\nu}$ of a $U(1)$ gauge field. In this case, and depending on the size of the coupling of the inflaton to the Chern-Simons term, there can be an explosive production of helical gauge fields at the end of inflation~\cite{Anber:2009ua,Anber:2015yca,Cado:2016kdp}. This exponential production can dominate the energy density of the Universe during the coherent oscillations of the inflaton around its minimum, a phenomenon dubbed as \textit{gauge preheating}~\cite{Adshead:2015pva,Sfakianakis:2018lzf,Cuissa:2018oiw}, and lead to a rapid production of inhomogeneities sourcing a significant gravitational wave background, leading to strong constraints on the inflaton Chern-Simons coupling from the Planck (and future CMB-S4) limits on the net energy density in gravitational waves~\cite{Adshead:2019lbr,Adshead:2019igv}. 

When the gauge field is identified with the SM hypercharge, $Y_\mu$, with strength $Y_{\mu\nu}$, the inflaton coupling to the Chern-Simons density $Y_{\mu\nu}\tilde Y^{\mu\nu}$ gives rise to the production of helical hypermagnetic fields which can then survive until the electroweak phase transition (crossover), and trigger the baryon asymmetry of the Universe (BAU)~\cite{Kamada:2016eeb,Kamada:2016cnb,Jimenez:2017cdr,Domcke:2019mnd,Cado:2021bia,Cado:2022evn}.
However, in the presence of strong gauge fields, light fermions charged under the gauge group are produced by the backreaction of gauge fields that source the fermion equations of motion (EoM)~\cite{Domcke:2018eki,Kitamoto:2021wzl}. The corresponding currents can then, in turn, backreact on the produced gauge fields, a phenomenon called \textit{Schwinger effect} (see e.g.~Ref.~\cite{Cohen:2008wz}). 
The backreaction of fermion currents on the produced gauge fields acts as a damping force in the explosive production of helical gauge fields, and many of the conclusions from the gauge field production should be revised in the presence of the Schwinger effect\footnote{One possible way out is if there are no light charged fields when gauge fields are produced, a condition that is not fulfilled here.}, in particular those concerning the preheating capabilities and the baryogenesis mechanism.

In this paper, we will study the effect of the Schwinger particle production on the helical hypermagnetic fields produced at the end of inflation, and in particular its influence on the gauge preheating efficiency and baryogenesis capability. In order to consider the backreaction of the produced gauge fields on the inflationary equations of motion and that of the Schwinger effect on the gauge field production, we will use numerical methods, in particular, the fourth order Runge-Kutta (\texttt{RK4}) algorithm. Our numerical results are validated as they overlap with some recent semianalytical methods and the gradient expansion formalism of Refs.~\cite{Sobol:2019xls,Gorbar:2021rlt,Gorbar:2021zlr,Fujita:2022fwc}.
Our general finding is that the gauge field production is much less explosive than in the absence of the Schwinger effect, which will jeopardize the conclusions concerning the possibility of gauge preheating, although they leave an open window for baryogenesis.

The contents of this paper are as follows. In Sec.~\ref{sec:model} we present the general lines of the model and the methods we will consider, including the relevant equations of motion in momentum space and the observable quantities we will compute. In Sec.~\ref{sec:slowroll} we will present numerical results for the gauge sector assuming the slow roll conditions in the inflaton equations of motion. 
In order to check the validity of our approach, numerical results will be compared with some estimates from the literature, namely the backreactionless solution, the Schwinger equilibrium and maximal estimates as well as the gradient expansion formalism where dynamical results are obtained analytically and numerically in configuration space. Some details about the numerical methods will be explained in App.~\ref{app:slow-roll}. In Sec.~\ref{sec:full-analysis} we will perform the full numerical calculation for two kinds of models that predict cosmological observables in agreement with the observed values: the $\alpha$-attractor models and the quartic hilltop models. In all cases the gauge preheating efficiency does not seem good enough to ensure complete reheating, which has to be completed by other perturbative or nonperturbative mechanisms. Moreover we have reanalyzed the baryogenesis predictions in the presence of the Schwinger effect and found a sizable window where the BAU is correctly predicted. 
Again, some details about the numerical methods we used are described in App.~\ref{app:full}. Finally, our conclusions are presented in Sec.~\ref{sec:conclusions}.

\section{The model}
\label{sec:model}

The model action is given by
\begin{align}
S= & \int d^4x \left[ \sqrt{-g} \left( \frac{1}{2}\partial_\mu \phi\partial^\mu \phi -\dfrac{1}{4} Y_{\mu\nu}Y^{\mu\nu} - V(\phi) \right)-\dfrac{\phi}{4f_\phi}\;Y_{\mu\nu}\tilde{Y}^{\mu\nu}  \right]
+\int d^4x \;\sqrt{-g}\; i\,\bar\psi \cancel{\mathcal D}  \psi ,
\label{eq:actionfinal}
\end{align}
where $\phi$ is the pseudoscalar inflaton, $V$ the inflaton potential, and $f_\phi$ provides the inverse coupling of the inflaton to the Chern-Simons term.  $Y^{\mu\nu}$ is the field strength of the hypercharge gauge field $Y^\mu$ and $\widetilde Y^{\mu\nu}=\frac{1}{2}\epsilon^{\mu\nu\rho\sigma}Y_{\rho\sigma}$ its dual tensor. 
 We also have included the interaction of fermionic currents, corresponding to hypercharge $Q_Y$ fermions, with the hypercharge fields (encoded in the covariant derivative $\mathcal D_\mu\equiv \partial_\mu-g' Q_Y A_\mu$).  All gauge field quantities are $U(1)$ hypercharge fields,  i.e.~$\bm{A}_Y$, $\bm{E}_Y$, $\bm{B}_Y$, etc. To make the notation lighter, we drop the index $Y$ as there will be no ordinary electromagnetic fields in this work.

 \subsection{Equations of motion}
 Variation of the action with respect to $\phi$ and the hypercharge gauge field $A_\mu= (A_0,\bm{A})$ leads to the gauge equations of motion in the radiation gauge, $A_0=0$ and $\bm{\nabla} \cdot \bm{A} =0$,
\begin{subequations} \label{def:EoM-system} \begin{eqnarray}
 &\ddot{\phi}+3H\dot{\phi}+V'(\phi)= \dfrac{\bm{E} \cdot \bm{B}}{a^4f_\phi}  ,\label{inflaton-EoM} \\&\left( \dfrac{\partial^2}{\partial \tau^2}-\nabla^2-\dfrac{a\, \dot{\phi}}{f_\phi}\; \bm{\nabla} \times \right) \bm{A}= \bm{J} ,\ese
 where we have used $Y_{\mu\nu}\tilde{Y}^{\mu\nu}=-4\, \bm{E} \cdot \bm{B}$ and $J^\mu=(\rho_c,\bm{J})=ig'Q_Y\bar{\psi}\gamma^\mu\psi$. 
We assume that initially the Universe does not contain any asymmetry of charged particles and that these ones are produced only later in particle-antiparticle pairs. Therefore, we set the charge density to zero, $\rho_c = 0$.
Finally, the current $\bm J$ is given by the Ohm's law
\be \bm{J} = \sigma \bm{E} = -  \sigma \dfrac{\partial \bm{A}}{\partial \tau},\label{Ohm-law}\ee
where $\sigma$ is the generalized conductivity, which will be defined later.

As it can be seen from the above system, we use cosmic time $t$ for the inflaton dependence and the conformal time $\tau$, defined by $g_{\mu\nu}=a^2(\tau)\,\eta_{\mu\nu}$, for the gauge field dependence. We denote the derivative with respect to conformal time $\tau$ with a prime and the derivative with respect to the cosmic time $t$ with a dot, e.g.~$a'=da/d\tau$ and $\dot{a}=da/dt$.
The Hubble parameter is defined as $H=\dot{a} (t)/a(t)$ where $a$ is the scale factor.
We assume a homogeneous inflaton with only zero mode, $\phi(t,\bm{x})=\phi(t)$.

We now quantize the gauge field $\bm{A}$ in momentum space
\be \bm{A}(\tau , \bm{x}) \, = \, \sum_{\lambda = \pm} \int \frac{d^3 k}{(2\pi)^3} \, \left [\bm{\epsilon}_\lambda(\bm{k}) \, a_\lambda(\bm{k}) \, A_\lambda(\tau, \bm{k}) \, e^{i \bm{k} \cdot \bm{x}} + \, \text{h.c.} \right]  ,
 \ee
where $\lambda$ is the photon polarization, and $a_\lambda(\bm{k})$ ($a_\lambda^\dagger(\bm{k})$) are annihilation (creation) operators that fulfill the canonical commutation relations
\be
[a_\lambda(\bm{k}),a_{\lambda'}^\dagger(\bm{k}')]=(2\pi)^3\delta_{\lambda\lambda'}\delta^{(3)}(\bm{k}-\bm{k'})\,.
\ee
The polarization vectors $\bm{\epsilon}_\lambda(\bm{k})$ satisfy the conditions\footnote{A simple realization can be given in terms of a real basis with the orthonormal vectors $(\bm{k}/|\bm{k}|, \bm{e}_i)$, ($i=1,2$), such that $\bm{k}\cdot \bm{e}_i=\bm{e}_1\cdot \bm{e}_2=0$ and $\bm{e}_i\cdot \bm{e}_i=1$, with $\bm{\epsilon}_\lambda\equiv(\bm{e}_1+i\lambda \bm{e}_2)/\sqrt{2}$, from where identities (\ref{eq:identitiespol}) follow.}
\be\begin{aligned} \bm{k} \cdot \bm{\epsilon}_\lambda(\bm{k}) &= 0\, , \hspace{2cm} &   \bm{k} \times \bm{\epsilon}_\lambda(\bm{k})  &= - i \lambda k \, \bm{\epsilon}_\lambda(\bm{k})\, , \\ \bm{\epsilon}^*_{\lambda'}(\bm{k}) \cdot \bm{\epsilon}_\lambda(\bm{k}) &= \delta_{\lambda \lambda'}\, , & \bm{\epsilon}^*_{\lambda}(\bm{k}) &= \bm{\epsilon}_\lambda(-\bm{k}) \, ,\end{aligned}
\label{eq:identitiespol}
\ee
where $k \equiv |\bm{k}|$.
Therefore, the equation of motion for the gauge modes yields
\be A''_\lambda+\sigma A'_\lambda  +k \left(k-\lambda \, \dfrac{a\, \dot{\phi}}{f_\phi}\right)A_\lambda =0\,.
\label{eq:Apm}
\ee
In some special cases ($\sigma=0$ and slow-roll inflation), this equation can be solved analytically, and we will do it in Sec.~\ref{sec:noSchwinger}. In the general case, we will solve it using numerical methods.

\subsection{Observable quantities}
Once we obtain a solution to the modes $A_\lambda$, we can compute the (hyper)electromagnetic energy densities as
\begin{subequations} \label{def:energy-densities} \begin{eqnarray}
\rho_{E} &\equiv& \frac{1}{a^4}  \int^{k_c}_{k_{\rm min}} dk \,  \frac{k^2}{4 \pi^2}\left(| A'_+|^2 + | A'_-|^2\right), \\
\rho_{B} &\equiv&\frac{1}{a^4} \int^{k_c}_{k_{\rm min}} d k \,  \frac{k^4}{4 \pi^2}\left(|A_+|^2 + |A_-|^2\right).\end{eqnarray} \end{subequations}
The upper integration limit comes because subhorizon modes have an oscillatory behavior and should be regarded as quantum fluctuations. Therefore, such modes do not contribute to the above classical observables and are excluded from the integration. More details and precise value of $ k_c$ will be given in Sec.~\ref{sec:BD-vacuum}. For the lower integration limit $k_{\rm min}$, see Eq.~(\ref{BD-penetration}).

In this work, we will also make use of the (hyper)magnetic helicity and its derivative, defined as
\begin{subequations} \label{def:AB-and-EB} \begin{eqnarray} \mathcal{H} &\equiv& \lim_{V\to\infty}\dfrac{1}{V}\int_Vd^3x \;\frac{\langle \bm{A} \cdot \bm{B} \rangle}{a^3} =\frac{1}{a^3}  \int^{k_c}_{k_{\rm min}}d k \, \frac{k^3 }{2\pi^2} \left(|A_+|^2-|A_-|^2\right), \label{ABdef} \\
\mathcal{G} &\equiv& \dfrac{1}{2a} \dfrac{d\mathcal{H} }{d\tau}  =-\lim_{V\to\infty}\dfrac{1}{V}\int_Vd^3x \;\frac{ \langle \bm{E} \cdot \bm{B}\rangle}{a^4}. \label{ABtoEB} \ese

In the case of one Dirac fermion with mass $m$ and hypercharge $Q_Y$, the conductivity can be written as\footnote{As the conductivity $\sigma$ relates $ \bm{J}$ and $\bm{E}$ in (\ref{Ohm-law}), it is a comoving quantity, i.e.~it scales with the Universe expansion. Our definition differs from the one in~\cite{Gorbar:2021rlt,Gorbar:2021zlr} where the authors used a physical conductivity that we will denote $\hat{\sigma}$ in this paper,  their relation being  $\sigma=a\,\hat{\sigma} $.\label{footnote:sigma}}~\cite{Domcke:2018eki}
\be
\sigma= \frac{|g'Q_Y|^3}{6\pi^2}
\frac{a}{H}\;\sqrt{2\rho_B}\;\coth\left(\pi\sqrt{\frac{\rho_B}{\rho_E}} \right)\exp{\left\{-\frac{\pi m^2}{\sqrt{2\rho_E}\, |g'Q_Y|}\right\}},
\label{eq:sigma-def}
\ee
where $g'\simeq0.4$ is computed at the characteristic scale $\mu\simeq (\langle \bm E\rangle^2+\langle \bm B\rangle^2)^{1/4}$ where the Schwinger effect takes place~\cite{Gorbar:2021zlr}.
This estimation is valid in the case of collinear electric and magnetic fields, an assumption that we have numerically checked by verifying that
\be
\cos\theta\equiv\frac{| \mathcal{G} |}{2\sqrt{\rho_{E} \rho_{B} }}\simeq 1\label{EM-angle},\ee
where $\theta$ is the spatial angle between $\bm{E}$ and $\bm{B}$. 

Moreover, the massless hypercharged fermions that are continuously produced during inflation have an energy density given by
\be \rho_\psi =  \lim_{V\to\infty}\dfrac{\sigma}{V}\int_Vd^3x \;\frac{\langle \bm{A} \cdot \bm{E} \rangle}{a^4} =\frac{\sigma}{a^4}  \int^{k_c}_{k_{\rm min}} d k \, \frac{k^2}{2\pi^2}  \dfrac{d}{d \tau} \left(|A_+|^2 +|A_-|^2  \right) \ee
Notice that the observable quantities $\rho_E$, $\rho_B$, $\rho_\psi$, $\mathcal H$ and $\mathcal G$ are physical\footnote{They relate to the comoving ones $\rho_E^c$, $\rho_B^c$, $\mathcal{H}^c$, and $\mathcal{G}^c$ by the relations
$ \rho_{B,E}^c = a^4 \rho_{B,E}$, $ \mathcal{H}^c=a^3\mathcal{H}$, $\mathcal{G}^c=a^4\mathcal{G}$.}, while the fields $\bm{A}$, $\bm{E}$ and $\bm{B}$ as well as the conductivity $\sigma$ and current $\bm{J}$ are comoving.

Concerning the Higgs vacuum expectation value, there are two possibilities during the inflationary period:
\begin{description}
\item[\textit{i)}]
The first possibility, which we will consider throughout this paper, is that $\langle h \rangle=0$, and so the electroweak symmetry is unbroken during the inflationary period.
In order to ensure unbroken electroweak symmetry and hence massless SM fermions, which all contribute to the conductivity (\ref{eq:sigma-def}), we assume that the SM Higgs field $h$ remains stabilized at the origin in field space by a large mass term throughout the inflationary period. Such a large mass can, e.g.,~be induced by a nonminimal coupling to the Ricci curvature scalar as $\mathcal L=\frac{1}{2}\xi h^2 R$ with $\xi>3/16$ (see e.g.~Ref.~\cite{Espinosa:2015qea}). Hence, we get
\be
\sigma\simeq\frac{41 \,g'^3}{72 \,\pi^2}\frac{a}{H}\;\sqrt{2\rho_B}\;\coth\left(\pi\sqrt{\frac{\rho_B}{\rho_E}} \right). \label{eq:sigma}
\ee
\item[\textit{ii)}]
The second possibility is that the electroweak symmetry is broken during the inflationary period. In this case after $\Delta N$ $e$-folds of inflation, there is a Gaussian distribution of values of the Higgs field with zero mean and variance 
$\langle h^2\rangle=H^2 \Delta N/(4\pi^2)$ with probability $P(h,\Delta N)\propto \exp(-\frac{1}{2}\frac{h^2}{\langle h^2\rangle})$ dominated by the values $h\lesssim \sqrt{\langle h^2\rangle}$, see Ref.~\cite{Espinosa:2015qea}\footnote{The SM Higgs potential is still unstable at a value of the Higgs field $h=h_I\simeq 10^{11}$ GeV and the condition for $P(h_I,\Delta N)<e^{-3 \Delta N}$ (so that it is unlikely to find the Higgs away from its EW vacuum in any of the $e^{3\Delta N}$ causally disconnected regions formed during inflation) implies $H_E<\sqrt{2/3}\,\pi h_I/\Delta N$, a condition that is not fulfilled by any of the models of inflation we have considered. Therefore this possibility would require stabilization of the Higgs potential by some new physics.}. In this case, the electroweak symmetry is broken and the hypercharge field strength $Y_{\mu\nu}$ in Eq.~(\ref{eq:actionfinal}) is projected onto the electromagnetic field strength $F_{\mu\nu}$ with a coupling to the inflaton given by $f_\phi/\cos^2\theta_W$ where $\theta_W$ is the electroweak angle. Now the conductivity for the hypermagnetic field in Eq.~(\ref{eq:sigma-def}) should be replaced by a similar expression for the magnetic field, with the replacement $|g' Q_Y|\to |e Q|$, where $e=gg'/\sqrt{g^2+g'^2}$ and $Q$ is the fermion electric charge. The condition for a fermion $f$ to contribute to the magnetic conductivity $\pi m_f^2< \sqrt{2\rho_E}\, |eQ_f|$ translates into the condition, for the fermion Yukawa coupling,
\be
Y_f\lesssim 0.45 \left(\frac{\rho_E}{H^4} \right)^{1/4} \sqrt{|Q_f|},
\ee
and we have computed all couplings at the characteristic scale $\mu\simeq (\langle \bm E\rangle^2+\langle \bm B\rangle^2)^{1/4}$ where the Schwinger effect takes place.
If the three generations of fermions satisfy the above condition then the conductivity for the magnetic field is given by Eq.~(\ref{eq:sigma}) with the replacement $\frac{41 g'^3}{72\pi^2} \to \frac{e^3}{\pi^2}$. We have checked that, in this case, the results for $f_\phi\lesssim 0.2\,\Mp$ are consistent with all three generation fermions contributing to the magnetic conductivity. For $f_\phi\gtrsim 0.2\,\Mp$ only the top quark does not contribute. Given that $41 g'^3/72\simeq 0.37$ while $e^3\simeq 0.36$, at the scales where the Schwinger effect takes place, we have found that the results in this second case are qualitatively similar to those for the previous case, which will be worked out in detail in this paper.

\end{description}

Considering then the case \textit{i)} above, the conductivity (\ref{eq:sigma})
yields a nontrivial integro-differential system as the damping term grows with the magnetic energy and hence backreacts on the amount of produced electric/magnetic fields.  
We aim to solve this setup of the Schwinger effect numerically. In the next sections we will consider specific cases where this system can be further simplified.

\subsection{The gauge vacuum \label{sec:BD-vacuum}}
At very early times, when $ |a \dot\phi | \ll k f_\phi$, the modes are in their Bunch-Davies (BD)  vacuum, hence 
\be A_\lambda(\tau,k) = \dfrac{1}{\sqrt{2k}} \;{\rm e}^{-i k \tau}\hspace{1.5cm} (\tau\to -\infty) . \label{vanilla-BD-vacuum}
\ee
Initially, we can consider all the modes in the BD vacuum (which would be possible by initializing the numerical simulation such that $a_0 \ll k_{0}/H_0$). In that case, since $|A_+| = |A_-|$, the fields $\bm{E}$ and $\bm{B}$ are plane waves perpendicular to each other, as $\mathcal{G}=0$ in (\ref{EM-angle}) yields $\cos\theta=0$. Therefore, there is no Schwinger effect and $\sigma =0$. 

It has recently been shown that in the presence of the conductivity $\sigma$, the BD vacuum amplitude of the modes that are still in the vacuum get damped by the ones that left it~\cite{Gorbar:2021rlt}.
Indeed, consider we are at a time $a_\ast$ where modes $k>k_\ast$ are still in the BD vacuum, while modes $k<k_\ast$ were amplified by both tachyonic and parametric instabilities from Eq.~(\ref{eq:Apm}).
Then, the equation of motion for modes such that $ |a_\ast \dot{\phi}(\tau_\ast) | \ll k f_\phi$ does not reduce to a plane wave in the presence of a non-zero $\sigma$, but instead to
$ A''_\lambda+\sigma A'_\lambda  +k^2 A_\lambda =0$,
and Eq.~(\ref{vanilla-BD-vacuum}) is not a solution anymore.
To derive the generalized BD vacuum, we write the gauge equation of motion (\ref{eq:Apm}) in cosmic time:
\be \ddot{A_\lambda}+\left(\hat{\sigma}+H\right)\dot{A_\lambda}+\frac{k}{ a}\left(\frac{k}{ a}-\lambda\frac{\dot{\phi}}{f_\phi}\right)A_\lambda=0, \ee
where we used the identity $a^{-2}A''_\lambda=\ddot{A}_\lambda +H\dot{A}_\lambda$,
and perform the transformation $A_\lambda= \sqrt{\Delta}\,\mathcal{A}_\lambda$ with~\cite{Gorbar:2021rlt}
\be \Delta(t) = \exp{\left\{-\int_{-\infty}^{t}\hat{\sigma}(t')\,dt'\right\}}.
\label{eq:Delta}
 \ee
We recall that we have defined $\hat{\sigma}=\sigma / a$ as the physical conductivity in  footnote~\ref{footnote:sigma}.
The above equation hence becomes
\be \mathcal{A}''_\lambda + \left[\frac{k}{ a}\left(\frac{k}{ a}-\lambda\frac{\dot{\phi}}{f_\phi}\right) -\frac{\dot{\hat{\sigma}}}{2}-\frac{\hat{\sigma}^2}{4}-\frac{H\hat{\sigma}}{2}\right] a^2 \mathcal{A}=0,\label{BD-vac-trans}\ee
where we used the fact that $\dot\Delta (t)=-\hat{\sigma}(t)\Delta (t)$.
A mode crosses the horizon when the expression in the square brackets vanishes for the first time at least for one polarization, at $k=k_c$. The modes in the vacuum are then characterized by $k\gg k_c$.
This yields the momentum of the mode that crosses the horizon at time $t$, namely the cutoff of the integrals:
\be k_c = \left|\frac{a\dot{\phi}}{2f_\phi} \right| + \sqrt{  \left(\frac{a\dot{\phi}}{2f_\phi} \right)^2+ \frac{a^2}{2}  \left[\dot{\hat{\sigma}}+\hat{\sigma}\left(\frac{\hat{\sigma}}{2}+H\right) \right]}\,. \label{cutoff-def} \ee

Deep inside the horizon, when the first term in square brackets of (\ref{BD-vac-trans}) dominates, the solution must satisfy the BD condition (\ref{vanilla-BD-vacuum}). As we have seen, in the presence of finite conductivity, this equation does not fully describe the gauge-field mode function inside the horizon, as the damped BD condition includes an exponential damping factor
\be A_\lambda (\tau, k) = \sqrt{\dfrac{\Delta (t)}{2k}} e^{-ik\tau}\hspace{1.5cm} (\tau\to -\infty).\ee

The bottom line of this section is that the modes still in their BD vacuum see their amplitudes damped because of the effect of the modes that left their vacuum earlier and participate in the equations of motion (\ref{inflaton-EoM}) and (\ref{eq:Apm}). The parameter $\Delta$ was first introduced in the context of the gradient expansion formalism in Ref.~\cite{Gorbar:2021rlt}, where it was dynamically solved, while in Ref.~\cite{Gorbar:2021zlr} it was also considered as a free parameter and validated the corresponding procedure by numerical calculations. In order to compare with results from the gradient expansion formalism in configuration space, we will also both compute $\Delta$ numerically and consider it as a free parameter, although our final results will be based upon the dynamical calculation of $\Delta$.

\section{Slow Roll Analysis}
\label{sec:slowroll}

The slow-roll inflation paradigm has been used by many authors to compute the amount of electromagnetic energy density~\cite{Anber:2006xt,Anber:2009ua,Durrer:2013pga}, or baryogenesis through helicity~\cite{Anber:2015yca,Cado:2016kdp,Domcke:2019mnd,Cado:2022evn} at the end of inflation, with or without taking into account the Schwinger effect. Here we aim to validate our numerical results by comparison with the known analytical results at the end of inflation. 

In this section we will take $\phi$ as a slowly rolling inflaton field such that 
$ \ddot{\phi} \simeq 0$, $3H \dot{\phi} \simeq - V'(\phi) $, and so 
we can consider $\dot{\phi}$ and $H=H_E$ as constant. Doing so, we are neglecting the gauge field backreaction in the right-hand side of Eq.~(\ref{inflaton-EoM}), a hypothesis that we have consistently checked \textit{a posteriori}.  The results of this section will be model independent, within the hypothesis of the slow roll approximation.

\subsection{Absence of Schwinger effect}
\label{sec:noSchwinger}
Here we are assuming there is no Schwinger effect\footnote{This condition should be considered as being fulfilled by some physical systems, as e.g.~systems with no massless fermions, more than as an approximation to the full (more realistic) case.}, i.e.~$\sigma=0$, hence we can rewrite (\ref{eq:Apm}) as 
\be A''_\lambda  +k \left(k+\lambda \, \dfrac{2 \xi}{\tau}\right)A_\lambda =0 ,
\label{eq:Apm-deSitter-nosigma}
\ee
where, following the slow roll equations,
\be \xi = -\dfrac{\dot{\phi}}{2H_E f_\phi} 
\label{eq:xi}
\ee
is a constant. Since we are in de Sitter space, we can use the scale factor definition $a=-(H \tau)^{-1}$ and solve (\ref{eq:Apm-deSitter-nosigma}) asymptotically. At early time, when $|k\tau|\gg 2\xi $, the modes are in their BD vacuum given by (\ref{vanilla-BD-vacuum}), as here $\Delta=1$.
When $|k\tau| \sim 2\xi $, one of the modes develops both parametric and tachyonic instabilities leading to exponential growth while the other stays in the vacuum. During the last $e$-folds of inflation, i.e.~$|k\tau|\ll 2\xi $, the growing mode has the solution~\cite{Anber:2006xt,Anber:2015yca}
\be A_\lambda \simeq \dfrac{1}{\sqrt{2k}}\left(\dfrac{k}{2\xi a_E H_E}\right)^\frac{1}{4} \exp{\left\{\pi \xi-2\sqrt{\dfrac{ 2\xi k}{a_E H_E}}\right\}}, \label{Amplified-mode} \ee
where $a_E$ and $H_E$ are, respectively, the scale factor and the Hubble parameter at the end of inflation. Here, as we assume a slow roll regime, we consider $H_E$ constant and we take the convention $a_E=1$.

Using (\ref{def:energy-densities}) and (\ref{def:AB-and-EB}) we can compute all electromagnetic quantities:
\begin{equation}
\rho_{E} \simeq  \dfrac{63}{2^{16}}\; \dfrac{H_E^4}{\pi^2\xi^3}\; {\rm e}^{2\pi \xi} ,  \quad
\rho_{B} \simeq  \dfrac{315}{2^{18}}\; \dfrac{H_E^4}{\pi^2\xi^5}\; {\rm e}^{2\pi \xi} ,  \quad
\mathcal{H}\simeq \dfrac{45}{2^{15}}\; \dfrac{H_E^3}{\pi^2\xi^4} \; {\rm e}^{2\pi \xi}, \quad
\mathcal{G} \simeq \dfrac{135}{2^{16}}\dfrac{H_E^4}{\pi^2\xi^4} \; {\rm e}^{2\pi \xi}.
\label{EBHel-deSitter-nosigma}
\end{equation}
These results are only valid when the absence of backreaction on the inflaton equation of motion (\ref{inflaton-EoM}) is guaranteed, hence when
$ \left| \mathcal{G}/V'(\phi)  \right|  \ll f_\phi$.
This model-dependent condition puts a lower bound on the parameter $f_\phi$ or, equivalently, a higher bound on $\xi$.
Using the slow roll equations and the definition of the slow roll parameters, this parameter can be written as 
\be \xi = \dfrac{\Mp}{ f_\phi}  \sqrt{\dfrac{\epsilon}{2}}, \label{xi-slowroll} \ee
where $\epsilon = (\Mp^2/2)\,  (V'/V)^2$. Therefore, at the end of inflation, where by definition $\epsilon=1$, one has $\xi=\Mp/\sqrt{2}f_\phi$, and the no backreaction condition in Eq.~(\ref{def:EoM-system}) provides the bound $\xi<5.73$ (or equivalently $f_\phi>0.12\, \Mp$).
In Fig.~\ref{fig:inflation-const-Delta} we show, with orange lines, the quantities $\rho_E$, $\rho_B$, $\mathcal H$ and $\mathcal G$ evaluated at the end of inflation obtained from the analytical backreactionless solutions from Eqs.~(\ref{EBHel-deSitter-nosigma}), while the blue dots are the numerical solutions, which correspond to the case $\sigma=0$ (no Schwinger effect) and correspondingly $\Delta=1$. We have used a Runge-Kutta method which is explained in App.~A.
\begin{figure}[htb]
\begin{center}
\includegraphics[width = 8.8cm]{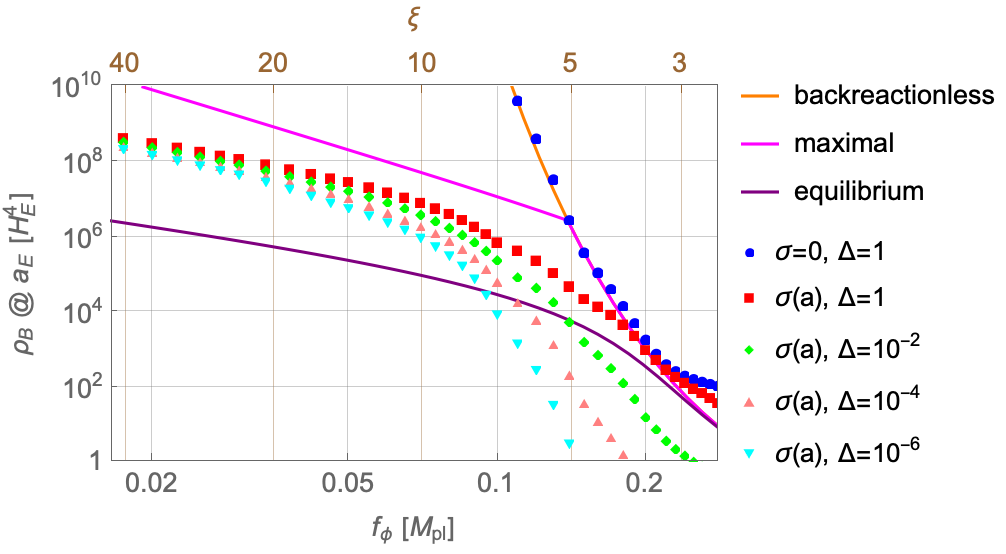}
\includegraphics[width = 6.2cm]{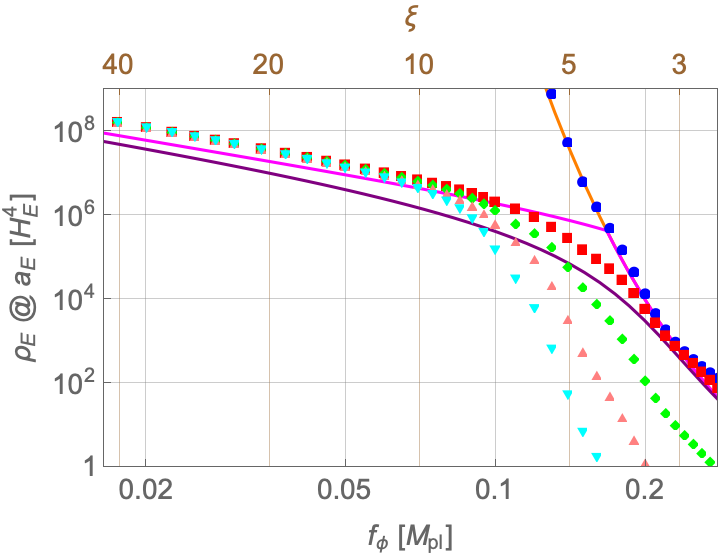}\\\vspace{4mm}
\includegraphics[width = 7.cm]{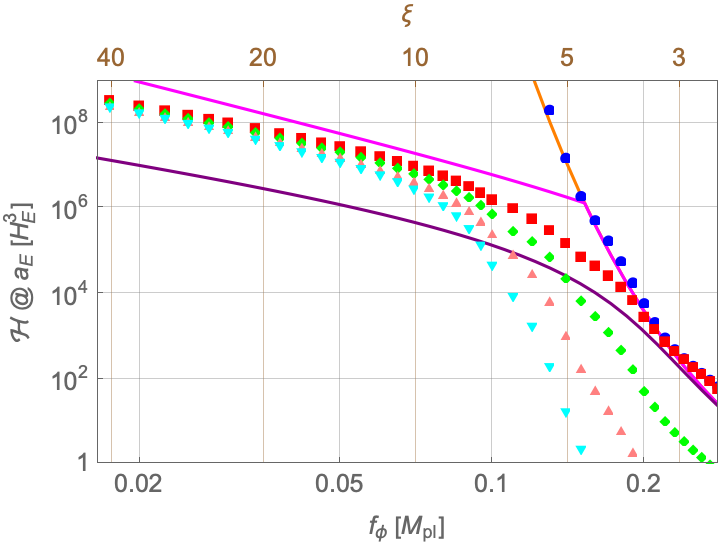}\hspace{10mm}
\includegraphics[width = 7.cm]{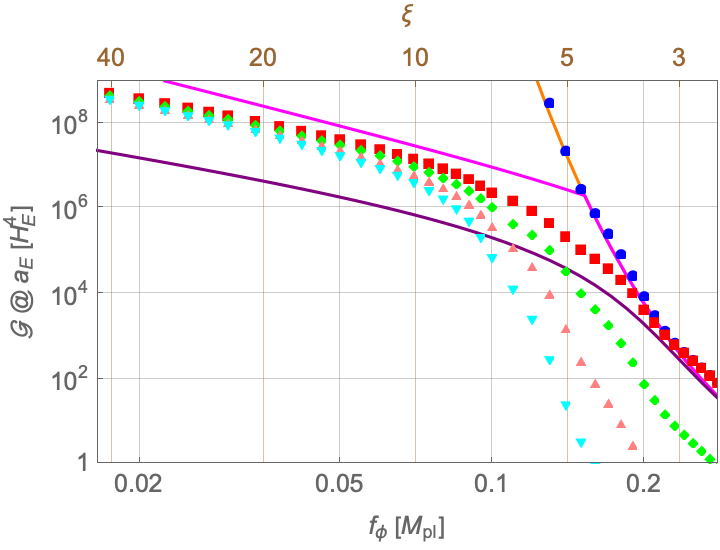}
 \caption{\it Electric $\rho_E$ and magnetic $\rho_B$ energy densities, and the helicity $\mathcal H$ and its derivative $\mathcal G$, at the end of inflation (i.e. for $\epsilon(a_E)=1$), in units of $H_E$, as functions of the coupling $f_\phi$ assuming $\Delta$ constant. We see the plots confirm the result from Fig.~1 of \cite{Gorbar:2021zlr}. Here, we also assumed $\xi$ constant.
 \label{fig:inflation-const-Delta}}
\end{center}
\end{figure}

\subsection{Presence of Schwinger effect}
\label{sec:Schwinger}
The Schwinger effect is taken into account by means of the conductivity $\sigma$
 in Eq.~(\ref{eq:Apm}), as given by Eq.~(\ref{eq:sigma-def})~\cite{Domcke:2018eki}. 
The growth of $\sigma$ with time then backreacts on the gauge field, as the damping term grows in its differential equation. We will compare our numerical calculations with three analytical (or semianalytical) results: the Schwinger maximal and equilibrium estimates~\cite{Domcke:2018eki,Gorbar:2021zlr}, as well as the gradient expansion formalism~\cite{Sobol:2019xls,Gorbar:2021rlt,Gorbar:2021zlr}. From the numerical point of view however, we aim to solve Eq.~(\ref{eq:Apm}) with $\sigma$ computed at each time step using (\ref{eq:sigma}). The details about the numerics will be displayed in Sec.~\ref{sec:slow-roll-num}.

\subsubsection*{Schwinger equilibrium estimate}
In this case, the backreaction of the chiral fermions on the gauge fields is taken into account by just replacing the parameter $\xi$ with the effective one~\cite{Domcke:2018eki}
\be \xi_{\rm eff} = \xi - \frac{41 \,g'^3}{144 \,\pi^2}\coth\left(\pi\sqrt{\frac{ \rho_{B}}{\rho_{E}}} \right)\frac{\sqrt{2\rho_E}}{H_E^2 }\,, \ee
in the backreactionless solutions (\ref{EBHel-deSitter-nosigma}). This amounts to solving
\be  \,\dfrac{63}{2^{15}\pi^2}\; \dfrac{{\rm e}^{2\pi \xi_{\rm eq}}}{\xi_{\rm eq}^3} =  \left( \frac{144\,\pi^2}{41 \,g'^3}\right)^2 (\xi -\xi_{\rm eq})^2 \;\tanh^2\left(\sqrt{\frac{5}{4}} \,\frac{\pi}{\xi_{\rm eq}} \right), \label{Schw-Equ-Xieff} 
\ee
which provides the function $\xi_{\rm eq}=\xi_{\rm eq}(\xi)$ that we plug in (\ref{EBHel-deSitter-nosigma}) instead of the bare $\xi$ to obtain the quantities $\rho_{\rm eq}^E$, $\rho_{\rm eq}^B$, $\mathcal H_{\rm eq}$ and $\mathcal G_{\rm eq}$. These equilibrium estimates are shown with a purple line in the plots of Fig.~\ref{fig:inflation-const-Delta}.

\subsubsection*{Schwinger maximal estimate}
In this case, we assume the exponential behaviors of the backreactionless solutions to be valid until they saturate the maximal value that we will display hereafter. We numerically determine the value of crossing, which happens for $\xi \simeq 4.4$-$4.7$ depending on each quantity.

The maximum helicity density can be estimated as the solution of~\cite{Domcke:2018eki}
\be |E|^2+|B|^2 = \xi_{\rm eff} |E|\,|B| .
\label{eq:restriccion}
\ee
This replacement yields an equation relating the $|E|$ and $|B|$ fields that can be solved analytically. We then choose, as definition of our maximal estimate, the solution $(|E|,|B|)$ of (\ref{eq:restriccion}) that maximizes the product $|E|\cdot|B|$~\footnote{Notice that our definition of maximal solution departs from that given in Refs.~\cite{Domcke:2018eki,Gorbar:2021zlr}, where the fields $|E|$ and $|B|$ are separately maximized, while we are maximizing the product $|E|\cdot|B|$, the relevant quantity for the baryon asymmetry generation.}. This yields for $\xi\gg 1$
\begin{subequations}\label{eq:EandBmax} \begin{eqnarray} \rho^E_{\rm max} &\simeq&   \frac{8}{9}\left(\frac{72\,\pi^2}{41\,g'^3} \right)^2\; \xi^2 H_E^4, \label{eq:Emax} \\
 \rho^B_{\rm max} &\simeq&   \frac{8}{81}\left(\frac{72\,\pi^2}{41\,g'^3} \right)^2 \;\xi^4 H_E^4, \label{eq:Bmax} \\
  \mathcal{H}_{\rm max} &=& \frac{2\,\mathcal{G}_{\rm max}}{3H_E}    \simeq \frac{32}{81}\left(\frac{72\,\pi^2}{41\,g'^3} \right)^2 \;\xi^3 H_E^3 .
 \ese

The maximal estimates for the quantities $\rho^E_{\rm max} $, $\rho^B_{\rm max} $, $\mathcal{H}_{\rm max} $ and $\mathcal{G}_{\rm max}$ are shown with a pink line in the plots of Fig.~\ref{fig:inflation-const-Delta}.
\subsubsection*{Gradient expansion formalism}
This method was introduced in Refs.~\cite{Sobol:2019xls,Gorbar:2021rlt,Gorbar:2021zlr} and transforms the EoM for the vector field $\bm A$ into EoM for observable quantities, in particular the electric $\bm E$ and magnetic $\bm B$ fields. As the spatial gradients in the EoM do always appear as ${\rm rot}\,\bm E $ and ${\rm rot}\,\bm B$, the EoM can be written as an infinite series in terms of the bilinears $\mathcal E^{(n)}=\langle \bm E\cdot \textrm{rot}^n \bm E\rangle/a^n$, $\mathcal G^{(n)}=\langle \bm E\cdot \textrm{rot}^n \bm B\rangle/a^n$ and $\mathcal B^{(n)}=\langle \bm B\cdot \textrm{rot}^n \bm B\rangle/a^n$, with $n=0,1, \dots$. In this way the coupled system of EoM for the fields $\bm E$ and $\bm B$ transforms into a system of coupled differential equations for the quantities $\mathcal E^{(n)}$, $\mathcal B^{(n)}$ and $\mathcal G^{(n)}$. This system is not block diagonal in the space of the $n$ index so that the system has to be truncated to find solutions.

Moreover, the parameter $\Delta(t)$ in Eq.~(\ref{eq:Delta}), which suppresses the gauge-field amplitude on small scales depends on the conductivity at all times $t'<t$. So, a precise determination of $\Delta(t)$ would require a complete analytical solution of the infinite-dimensional system of equations. 
While $\Delta$ was dynamically computed in Ref.~\cite{Gorbar:2021rlt}, for the sake of simplicity and generality, it was considered as a free parameter in Ref.~\cite{Gorbar:2021zlr} and fixed to the values $\Delta=1,\, 10^{-2},\, 10^{-4},\, 10^{-6}$. In our numerical approach we will consider $\Delta$ as a function of the conductivity $\sigma$, as the initial condition for $\bm E$ and $\bm B$ are plane waves, such that $\bm E\cdot \bm B=0$ and therefore initially $\sigma=0$ and so $\Delta=1$. However, as time is evolving $\bm E$ and $\bm B$ will become collinear, and a nonvanishing conductivity will appear, as well as the function $\Delta(t)<1$. In order to compare our numerical results with those from Ref.~\cite{Gorbar:2021zlr}, we also will eventually enforce $\Delta$ to be a constant in our code. Upon considering a constant value of $\Delta$, our results will agree pretty well with those obtained in the gradient expansion formalism, see Fig.~\ref{fig:inflation-const-Delta}. In the more realistic cases where we just compute the value of $\Delta(t)$, we will see that at the beginning $t=t_0$, just very deep inside the inflationary period, $\Delta(t_0)=1$, while the value of $\Delta$ will decrease very fast and at the end of inflation $t=t_E$, $\Delta(t_E)\ll 1$. 

\subsection{Numerical results at the end of inflation\label{sec:slow-roll-num}}
We will find it more convenient to change the variable from the time $t$ to the scale factor $a$.
The gauge field equation of motion (\ref{eq:Apm}) then becomes 
\be \frac{\partial^2 A_\lambda}{\partial a^2}+ \frac{1}{a}\left(2+\frac{\sigma}{aH_E}\right)\frac{\partial A_\lambda}{\partial a} +\frac{k}{a^3H_E}\left(\frac{k}{aH_E}-2\lambda \xi \right)A_\lambda=0.
\label{eq:Apm-vs-a}
\ee
We recall that, as we are considering the slow roll regime in this section, we do not need to solve the equation of motion for $\phi$.

The Bunch-Davis solutions can now be written as
 \be \ba A_\lambda (a,k) &= \sqrt{\frac{\Delta(a)}{2k}}\; {\rm e}^{i k/aH_E} \\
\frac{\partial A_\lambda}{\partial a} (a,k)&= \frac{\sqrt{\Delta(a)}}{a^2H_E}\left( -i\,\sqrt{\dfrac{k}{2}}-\frac{\sigma}{2}\dfrac{1}{\sqrt{2k}}  \right){\rm e}^{i k/aH_E} \ea \hspace{1.2cm} (a\to0), \label{BD-modes-a} \ee
with
\be \Delta(a) = \exp{\left\{-\int_{a_0}^{a}\frac{\sigma(a')}{a'^2 H_E}da'\right\}}. \label{Delta-of-a}\ee
The technical details of the numerical simulations for solving Eq.~(\ref{eq:Apm-vs-a}), subject to the boundary conditions (\ref{BD-modes-a}), can be found in App.~\ref{app:slow-roll}.
We display in Fig.~\ref{fig:spectra-many-a} the spectra of all the observable quantities in order to see how the BD vacuum is dominating the spectra for large $k$ and how the cutoff $k_c(a)$, given by (\ref{cutoff-def}), efficiently removes that part of the integration. The difference between the BD vacuum and the damped BD vacuum is also clear, as the first goes like $k^3$ whereas the second goes like $\Delta(a)k^3$ with $\Delta$ decreasing with time. Hence the asymptotic behaviors are not superimposed since $\Delta$ changes. Finally, we also see explicitly how the growth of $\rho_E$ and $\rho_B$ with the scale factor $a$ is due to the increase in amplitude of the spectrum hump and its shift to larger values of $k$. For this illustrative purpose we used a constant value of $\xi$. Here we have fixed $f_\phi =0.1\, \Mp$, while for other values of this parameter the plots are similar.

Before moving to the full numerical results, we will compare our slow roll based inflaton numerical results with the recent literature on the subject. 
\begin{figure}[b]
\begin{center}
\includegraphics[width = 8.55cm]{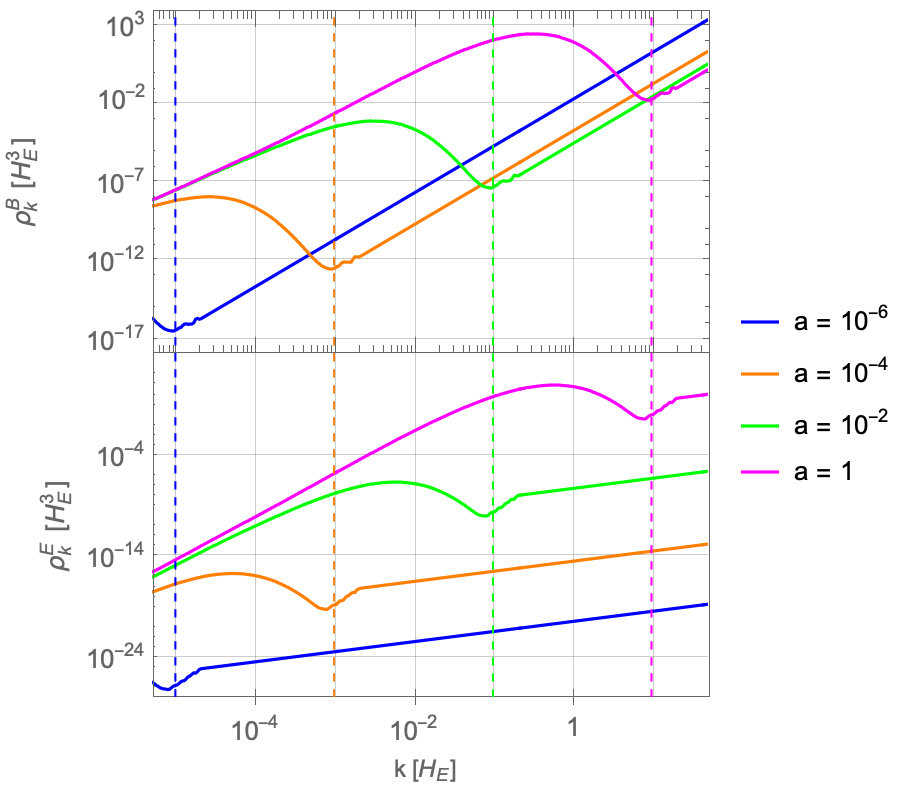}
\includegraphics[width = 6.45cm]{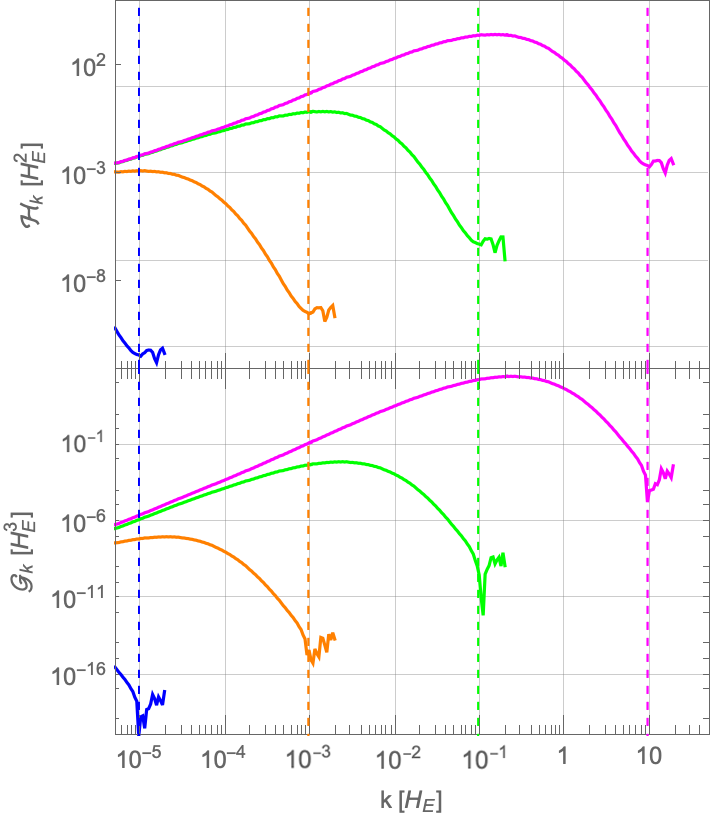}
 \caption{\it Spectra of the magnetic energy (top left), electric energy (bottom left), helicity (top right) and its derivative (bottom right), i.e.~the integrands of (\ref{def:energy-densities}) and (\ref{def:AB-and-EB}), for different values of $a$ during inflation simulation. Here we used variable $\sigma(a)$ and $\Delta(a)$ with constant $\xi$. The color matching dashed vertical lines show the cutoff values $k_c (a)$ computed from (\ref{cutoff-def}).
  \label{fig:spectra-many-a}}
\end{center}
\end{figure}

\subsubsection{Constant $\Delta$ and $\xi$ approximation}
We will first assume that the parameters $\Delta$ and $\xi$ are constants. As we already mentioned, the parameter $\Delta$ was fixed to constant values in Ref.~\cite{Gorbar:2021zlr} while $\xi$, as defined in Eq.~(\ref{eq:xi}), is often considered as a constant in the slow roll approximation. 
In Fig.~\ref{fig:inflation-const-Delta}, we displayed several results already present in the literature that we successfully reproduced with our numerical method. First the backreactionless case, where there is no conductivity, by simply enforcing $\sigma=0$ (therefore $\Delta=1$) in the code. The data set are displayed in blue and match the corresponding analytical value given by Eq.~(\ref{EBHel-deSitter-nosigma}). Then, in order to reproduce results from~\cite{Gorbar:2021zlr}, we considered a non-zero conductivity given by (\ref{eq:sigma}) while assuming $\Delta$ constant during inflation, thus making it a free parameter. In Fig.~\ref{fig:inflation-const-Delta}, we plot the quantities $\rho_B$, $\rho_E$, $\mathcal{H}$ and $\mathcal{G}$ at the end of inflation for chosen values of $\Delta$.  We can see that the results agree well with those using the gradient expansion formalism in Ref.~\cite{Gorbar:2021zlr}.

\subsubsection{Variable $\Delta$ and $\xi$}
The benefit of the slow roll approximation is that the results look ``model independent". However, the tradeoff comes with the need of having a constant parameter $\xi$ as the slow roll regime implies an approximately constant $\dot{\phi}$. Besides, we know that this parameter can also be expressed in terms of the slow roll parameter $\epsilon$ (see Eq.~(\ref{xi-slowroll})), which is indeed small and constant during inflation but then quickly becomes unity during the last $e$-folds. We also know that the modes produced during the last $e$-folds are the ones that contribute the most to the integrals (\ref{def:energy-densities}) and (\ref{def:AB-and-EB}), as all the modes previously generated get washed out by the Universe expansion.

All these observations lead us to conclude that the most important contribution to the quantities $\rho_E$, $\rho_B$, $\mathcal H$ and $\mathcal G$ is taking place during an epoch when the constant $\xi$ approximation loses its relevance. Hence, in this section, we will instead specify an inflation model, namely the Starobinsky potential, and make its study in the slow roll regime with a function $\xi(a)$ that can be obtained from the model. We have chosen in this section the Starobinsky potential as it provides a realistic model of inflation, and will be a particular case of a more general class of models we will consider to make predictions using the full solution of the system. The purpose of this section will thus be to assess the goodness of the slow roll approximation when computing the full solution to the system (\ref{def:EoM-system}).

The Starobinsky potential is given by
\be
V = \Lambda^4 \left[ 1- \exp{\left\{-\sqrt{\dfrac{2}{3}} \dfrac{|\phi|}{\Mp}\right\}} \right]^2 . \label{eq:Starobinsky-potential}
\ee
Using the slow roll regime, the inflaton field $\phi$ is given by
\be
\sqrt{\frac{2}{3}}\,\frac{\phi(a)}{\Mp}=-\log\left(\dfrac{a_E}{a}\right)^{\frac{4}{3}}-\mathcal W_{-1}\left[-\beta e^{-\beta}\left(\dfrac{a_E}{a}\right)^{-\frac{4}{3}}\right]-\beta+\log\beta,\quad \beta=1+\dfrac{2}{\sqrt{3}}
\ee
where $\mathcal W_n$ is the $n$th branch of the Lambert function.
The value of the function $\xi$ is then given by
\be
\xi(a)=\sqrt{\frac{2}{3}} \frac{\Mp}{f_\phi}\frac{1}{\exp\left[\sqrt{\frac{2}{3}}\frac{\phi(a)}{\Mp}\right]-1}\,.
\ee

In Fig.~\ref{fig:inflaton-result-fphi}, we display in blue results for the Starobinsky model,  for various values of $\epsilon$, when $\sigma$ and $\Delta$ vary dynamically. Although the slow roll approximation loses its relevance for values of $\epsilon$ closer to 1 (an issue we address in the next section), we already see a difference with the plots in Fig.~\ref{fig:inflation-const-Delta}. This is because, no matter the value of the initial time, the function $\Delta (a)$ rapidly goes to extremely small values, thus killing the BD modes that would have been amplified at the very end of inflation and that would have contributed the most to the integrals (\ref{def:energy-densities}) and (\ref{def:AB-and-EB}). With a constant $\Delta$, this suppression is less effective and the tachyonic amplification yields higher energy densities and helicity. 
\begin{figure}[htb]
\begin{center}
\includegraphics[width = 8.9cm]{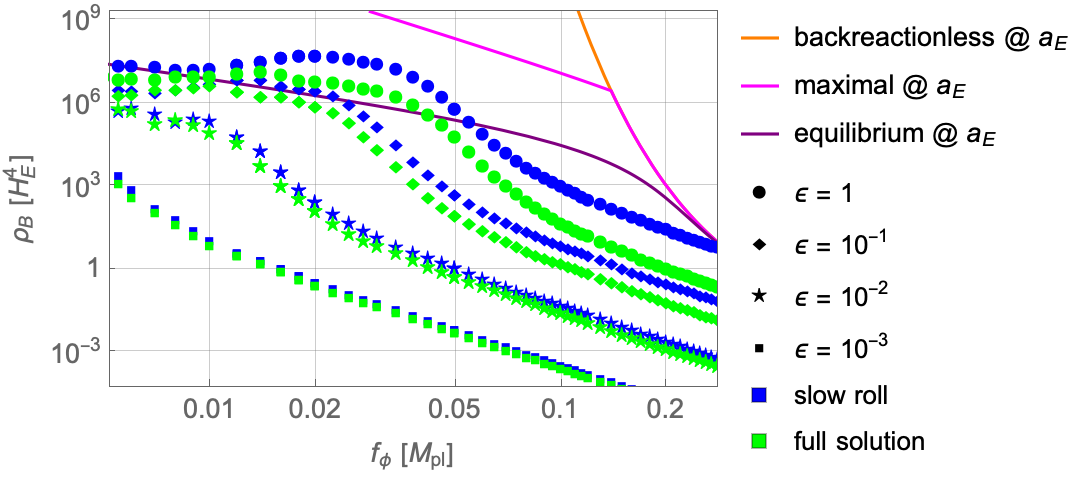}
\includegraphics[width = 6.1cm]{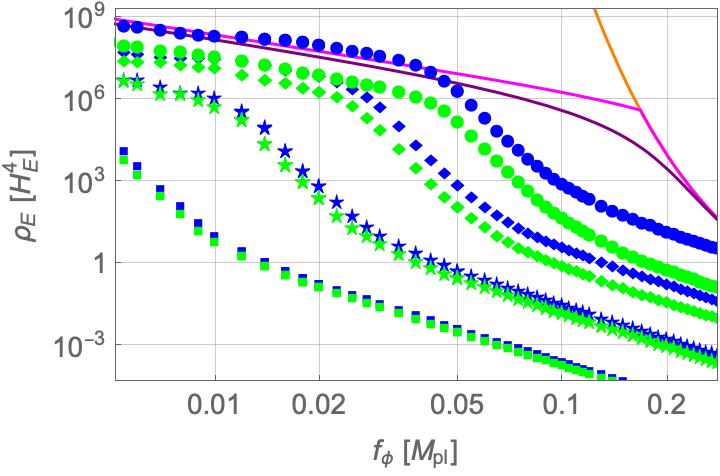}\\\vspace{3mm}
\includegraphics[width = 6.8cm]{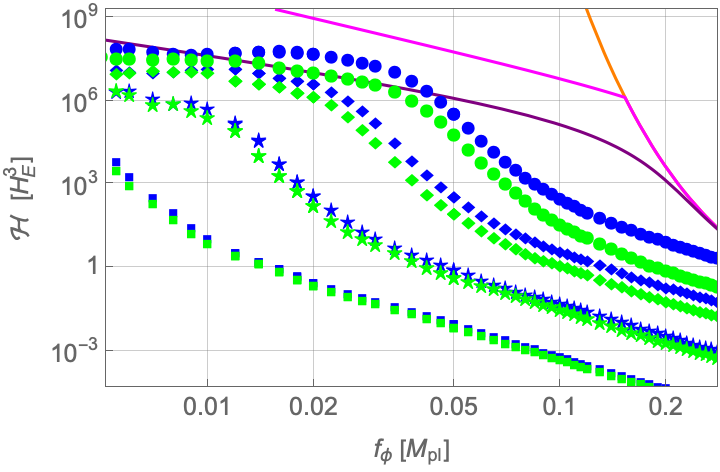}\hspace{14mm}
\includegraphics[width = 6.8cm]{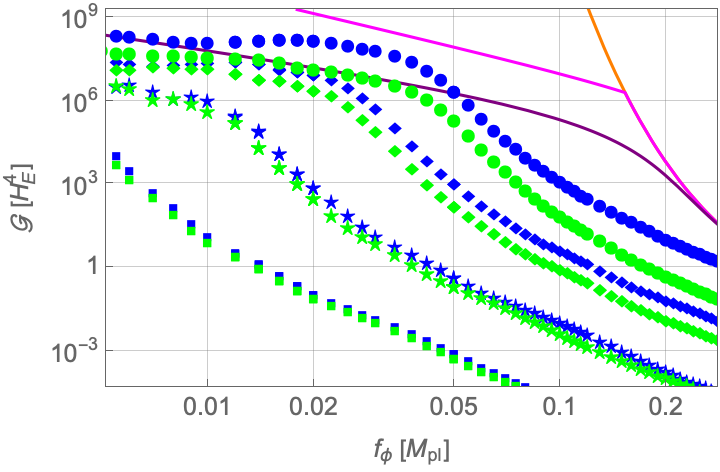}
 \caption{\it Comparison between the slow roll approximation and the full solution for the Starobinsky model. The analytical estimates are given for $\epsilon=1$. As expected, the slow roll computation diverges from the full solution as inflation is nearing the end, since the slow roll approximation is only valid in the regime $\epsilon\ll1$. Hence the slow roll computation overshoots the value of all quantities, closer to the value given by the Schwinger equilibrium estimate for $f_\phi \lesssim 0.05\, \Mp$. As expected, we also have compared both analysis, slow roll and full solution, for values of $a$ such that $\epsilon(a)\ll1$ (in particular $\epsilon=10^{-1},\, 10^{-2},\, 10^{-3}$) and found good agreement.
 \label{fig:inflaton-result-fphi}}
\end{center}
\end{figure}

\section{Full Analysis}
\label{sec:full-analysis}

In this section, we are not using the slow roll hypothesis for the inflaton equation of motion and consider the full solution to the system (\ref{def:EoM-system}) in specific models of inflation. We will choose a set of inflationary models that are well known to be in agreement with all cosmological constraints. Also, we do not assume any peculiar geometry of the Universe.

The equations to be solved during inflation are the system (\ref{def:EoM-system}) written in terms of the variable $a$. Unlike in the previous section, the current change of variables must take into account that the Hubble parameter is not constant, but moreover we have 
$ \frac{da}{dt}= \dot{a}= aH$, and we will define the auxiliary quantity $\mathcal F$ as
\be \mathcal{F} =- \frac{a}{H}\frac{dH}{da} =- \frac{a}{2H^2}\frac{dH^2}{da}. \ee
We will relate it to the Friedmann equations
\bse H^2 &=& \frac{\rho}{3 \Mp^2}, \label{eq:Friedman1} \\ \frac{\ddot a}{a}&=&-\frac{3p+\rho}{6M_p^2}, \label{eq:Friedman2}\ese
which combine themselves into
\be \frac{a}{2}\frac{dH^2}{da}= \frac{dH}{dt}= \frac{\ddot a}{a} - H^2 = -\frac{p+\rho}{2M_p^2}, \ee
where the total energy density and pressure are
\bse \rho &=& \frac{1}{2}\dot\phi^2+V+\rho_{\rm EM} + \rho_\psi, \label{eq:total-rho}\\  p &=& \frac{1}{2}\dot\phi^2-V+\frac{\rho_{\rm EM} }{3}+\frac{\rho_\psi }{3}. \ese
Hence we have
\be H^2 \mathcal{F} =-\frac{a}{2}\frac{dH^2}{da}=\frac{1}{\Mp^2}\left(\frac{1}{2}\dot\phi^2+\frac{2}{3}\rho_{\rm EM}+\frac{2}{3}\rho_\psi\right). \ee
and the system (\ref{def:EoM-system}) becomes 
\begin{subequations}\label{Staro-EoMs} \begin{eqnarray}  \frac{d^2\phi}{da^2}+\frac{4  -\mathcal{F}}{a}\frac{d\phi}{da}&+&\frac{V'(\phi)}{a^2H^2}+ \frac{\mathcal{G}}{a^2H^2 f_\phi}=0 \\
\frac{d^2 A_\lambda}{d a^2}+\frac{1}{a}\left(2 -\mathcal{F}+\frac{\sigma}{aH}\right)\frac{d A_\lambda}{d a} &+&\frac{k}{a^2H}\left(\frac{k}{a^2H}-\frac{\lambda }{f_\phi} \frac{d\phi}{da} \right)A_\lambda=0 . \ese

The Hubble parameter can be computed from the Friedmann equation (\ref{eq:Friedman1}), where $\rho$ is given by (\ref{eq:total-rho}).
This way, we can compute the value of $H$ and $\mathcal{F}$ at each time step recursively to feed the equations of motion, like we already do for $\sigma$ and $\mathcal{G}$. The BD vacuum modes are identical to the previous case, see Eqs.~(\ref{BD-modes-a}). Finally, for comparison purposes, we can define a generalized time dependent instability parameter $\xi(a)$ as
\be \xi(a) = -\frac{a}{2f_\phi}\frac{d\phi}{da} \label{eq:xi-a}\ee
such that it corresponds to the definition (\ref{eq:xi}).
The simulations show that this parameter, obtained from full solution computation, significantly differs from the slow roll one at the very end of inflation.

\subsection{Full numerical results at the end of inflation}
\label{sec:numerical-end}

In this subsection we will compare our results at the end of inflation, where we are making a full numerical analysis of the EoM, with those obtained using the slow roll approximation for the inflationary potential. For the sake of comparison we will concentrate on the Starobinsky model given by (\ref{eq:Starobinsky-potential}). In this current framework, we see in Fig.~\ref{fig:inflaton-result-fphi} that the four studied quantities, namely $\rho_B$, $\rho_E$, $\mathcal{H}$ and $\mathcal{G}$, are much closer to the Schwinger equilibrium estimate at the end of inflation.

We present in Fig.~\ref{fig:inflaton-result-fphi} the values of the physical observables evaluated at various stages of inflation, i.e.~various values of the scale factor $a$, from $\epsilon(a) = 10^{-3}$ to  $\epsilon(a)=1$, as a function of the coupling $f_\phi$ for the Starobinsky model. We superimpose the analytical results from Secs. \ref{sec:noSchwinger} and \ref{sec:Schwinger}, and hence the backreactionless solution as well as the Schwinger maximal and equilibrium estimates. From the plots we see that for $f_\phi\lesssim0.05\,\Mp$ the equilibrium estimate is a good approximation, especially for $\rho_E$ where the predictions of maximal and equilibrium estimates merge.
We also verify that $\cos \theta \simeq 1$ hence satisfying the assumption on parallel electric and magnetic fields leading to the conductivity definition (\ref{eq:sigma-def}).

In this setup, our numerical code is computing a value of the conductivity $\sigma$ and $\Delta$ for each time step, hence we got the functions $\sigma(a)$ and $\Delta(a)$.
The variation and presence of $\Delta(a)$ is not without effect on the final results. Indeed, the smallest ($k\gg H_E$) modes are the ones that most contribute to the integrals (\ref{def:energy-densities}) and (\ref{def:AB-and-EB}). Without the Schwinger effect, these modes are produced last, just at the end of inflation, and only briefly leave the horizon. They therefore should have a significant impact on preheating. When the Schwinger effect prevents their generation, by reducing them by a $\ll 1$ factor, while they are still in the BD vacuum, we can ask ourselves about the effectiveness of gauge preheating.
It was shown in previous studies of gauge preheating~\cite{Cuissa:2018oiw} that its efficiency mainly depends on the electromagnetic energy fraction available at the end of inflation $\rho_{\rm EM} / \rho_{\rm tot}$. To shed light on the last point, we will extend, in the next section, our numerical results beyond the end of inflation when the inflaton is coherently oscillationg around its potential minimum. We will do that in a set of particularly interesting phenomenological models that we describe in the next section.

\subsection{Inflationary models}
We will here introduce two classes of models that all satisfy the cosmological constraints. They should be considered as a sample of possible models, and they are just chosen for illustrative purposes, as they do not exhaust by any means the allowed inflationary models. 

\subsubsection{$\alpha$-attractor models}
The $\alpha$-attractor potential is given by~\cite{Kallosh:2013yoa}
\be V_\alpha(\phi) = \Lambda_\alpha^4 \left[ 1- \exp{\left\{-\sqrt{\dfrac{2}{3\alpha}} \dfrac{|\phi|}{\Mp}\right\}} \right]^2 .\ee
Setting $\alpha=1$ yields the $R^2$ model or Starobinsky potential (\ref{eq:Starobinsky-potential}). To make the comparison interesting, we choose to have $1\leq\alpha \leq 100$, where cosmological observables are correctly reproduced. In the slow roll approximation, the field value at the end of inflation is
\be \phi_E = \sqrt{\dfrac{3 \alpha}{2}} \; \Mp \, \log{\left(1+\dfrac{2}{\sqrt{3\alpha}}\right)}. \ee

We can readily compute $\phi_\ast$, and evaluate the slow roll parameters $N_\ast=60$ $e$-folds before the end of inflation. The slow roll parameters and the cosmic observables are in agreement with the cosmological contraints for the range
\be 1 \leqslant \alpha \lesssim 100. \ee
In particular, for $\alpha = 1\, (100)$ we get
\be \ba \epsilon_\ast &\simeq 0.00019 \ (0.00387), \hspace{2cm} \eta_\ast \simeq -0.0159\ (-0.00331) \\ n_s &\simeq0.967 \ (0.97), \hspace{1.cm} r_\ast \simeq 0.003\ (0.062),\hspace{1.cm}  H_E \simeq 0.82 \ (1.13)\, \cdot 10^{13} \ \text{GeV}. 
\ea \ee
in agreement with the observed values~\cite{Planck:2018jri}
\be
n_s^{\rm obs}\simeq 0.9649\pm 0.0042,\quad r_\ast^{\rm obs}\lesssim 0.06,\quad H_\ast^{\rm obs}\lesssim
6\cdot 10^{13}\, \textrm{GeV}\ (95\%\,\textrm{CL}).
\ee

Using the observed value of $A_s$ from Ref.~\cite{Planck:2018jri}, $A_s^{\rm obs}=2.2\cdot10^{-9}$,  we fix the vacuum energy. The result depends on $\alpha$ and is approximately given by $\Lambda_\alpha \simeq 3.4 \cdot 10^{-3}\,  \alpha^{1/5} \,\Mp$. We then obtain the values $\Lambda_{1} = 3.152\cdot 10^{-3}\,\Mp$ and $\Lambda_{100} = 8.313\cdot 10^{-3}\,\Mp$.
\begin{figure}[htb]
\begin{center}
\includegraphics[width = 7.7cm]{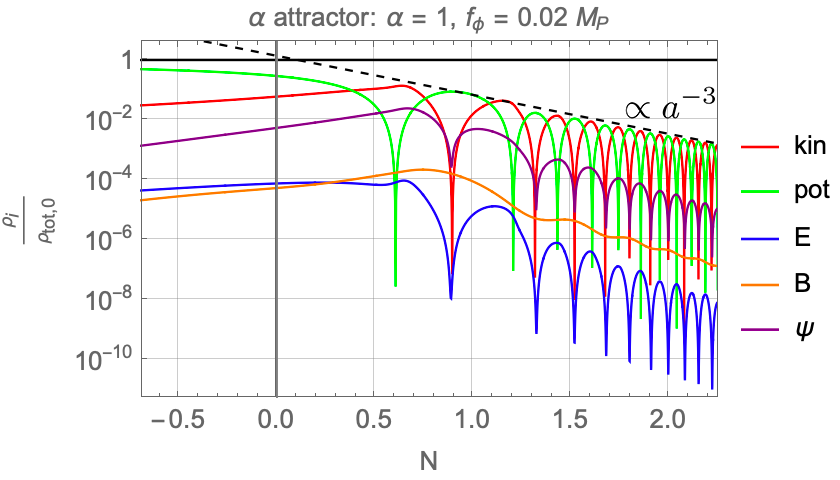} 
\includegraphics[width = 6.8cm]{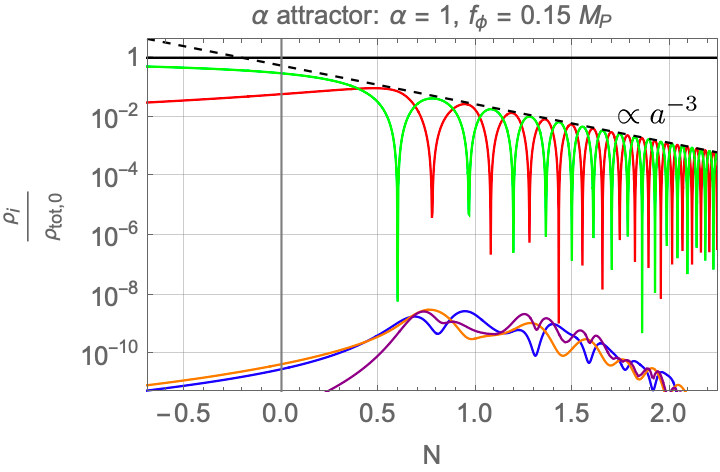}
\\\vspace{2mm}
\includegraphics[width = 7cm]{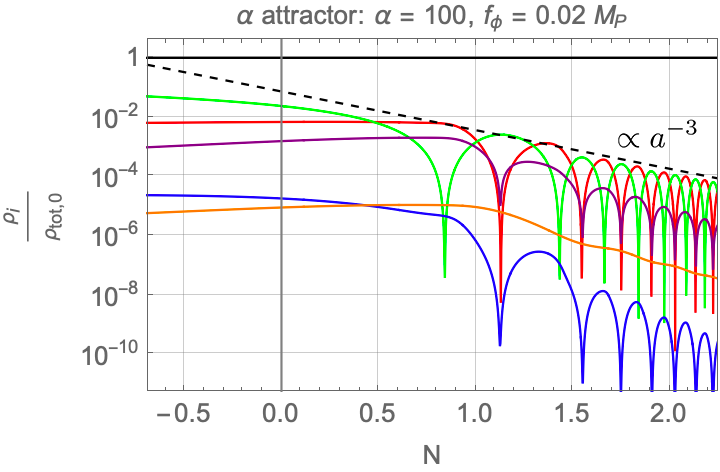}\hspace{5mm}
\includegraphics[width = 7cm]{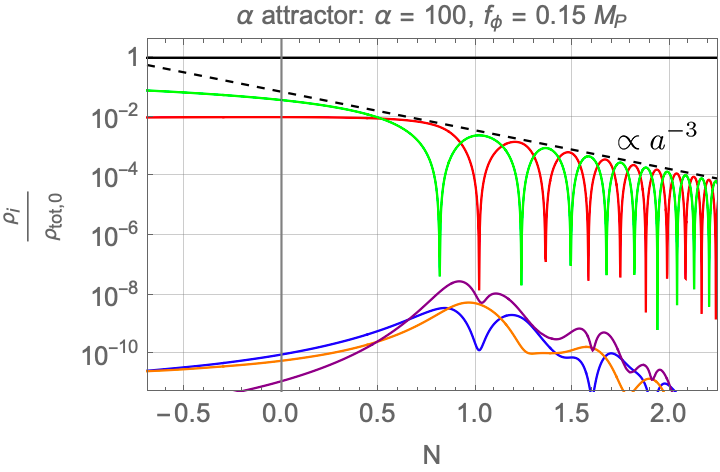} 
 \caption{\it Inflaton kinetic and potential energy density, as well as electric, magnetic and fermion energy density ratios to the initial total energy density of the Universe for the $\alpha$-attractor models with $\alpha=1$ (upper panels) and $\alpha=100$ (lower panels). The vertical gray lines display the value $a$ for which $\epsilon(a)=1$ and the dashed line shows the expected scaling of the dominant sector.
 \label{fig:Preheating-energy-breakdown-vA}}
\end{center}
\end{figure}

\subsubsection{Hilltop quartic models}

The hilltop model potential is given by~\cite{Boubekeur:2005zm}
\be V_{\rm h}(\phi) = \Lambda_{\rm h}^4 \left[ 1-  \left(\dfrac{\phi}{\mu}\right)^p\right]^2.\ee
The case $p = 4$ can be compatible with the Planck measurements. There are two ways for the field to relax to the minimum at $\phi=\mu$, with different initial conditions:
\begin{enumerate}
\item $\phi_\ast > \phi_E$: In this case the field $\phi> \mu$ is relaxing in a potential region that can be approximated by $V_{\rm h}\sim \phi^8$, and thus, the slow roll conditions are not met, as chaotic inflation is ruled out.
\item $\phi_\ast <\phi_E$: In this case the field $\phi< \mu$ is relaxing in a flat potential region and the model predicts correct inflationary observables for a large range of the parameter. In this work, we will study this option.
\end{enumerate}

The slow roll parameters and the cosmic observables are in agreement with the contraints for the range
\be 10 \Mp\lesssim \mu \lesssim 50\Mp. \ee
We fix the vacuum energy from the constraint on the amplitude of scalar fluctuations. The result depends on $\mu$ and is approximately $\Lambda_{\rm h} \simeq 6\cdot 10^{-4}\, \mu^{2/3} \,\Mp^{1/3}$. We then have the values $\Lambda_{\rm h} = 3.243\cdot 10^{-3}\,\Mp$ for $\mu=10\,\Mp$ and $\Lambda_{\rm h} = 8.081\cdot 10^{-3}\,\Mp$ for $\mu=50\,\Mp$.

In particular, for $\mu = 10 \ (50)\,\Mp$ we get
\be \ba \epsilon_\ast &\simeq 0.00021 \ (0.0041) \hspace{2.4cm} \eta_\ast \simeq -0.0207\ (-0.00328) \\ n_s &\simeq0.957 \ (0.97) \hspace{1.3cm} r_\ast \simeq 0.00335\ (0.0654),\hspace{1.3cm} H_E \simeq 0.64 \ (1.1)\, \cdot 10^{13} \ \text{GeV} . \ea \ee

\begin{figure}[htb]
\begin{center}
\includegraphics[width = 7.7cm]{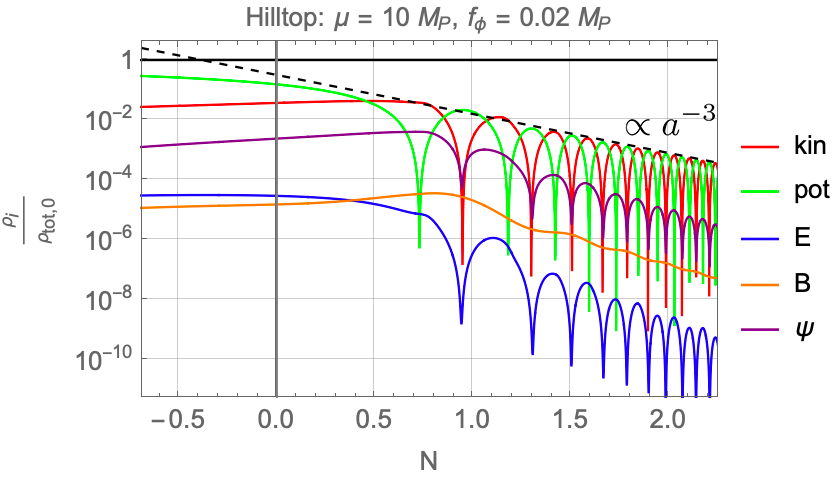} \hspace{2mm}
\includegraphics[width = 6.8cm]{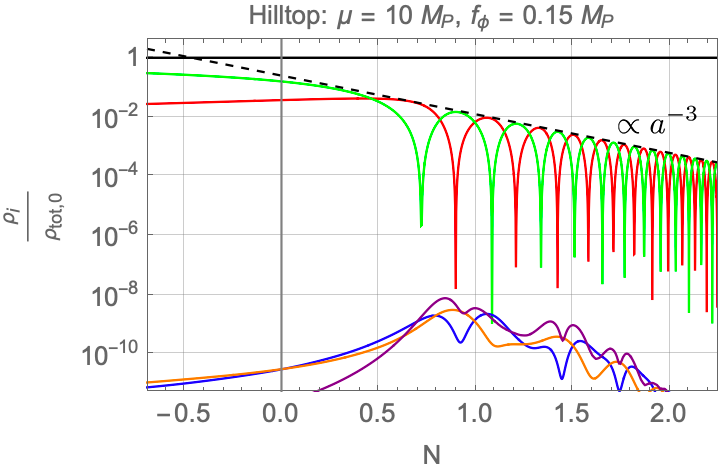}
\\\vspace{2mm}
\includegraphics[width = 7cm]{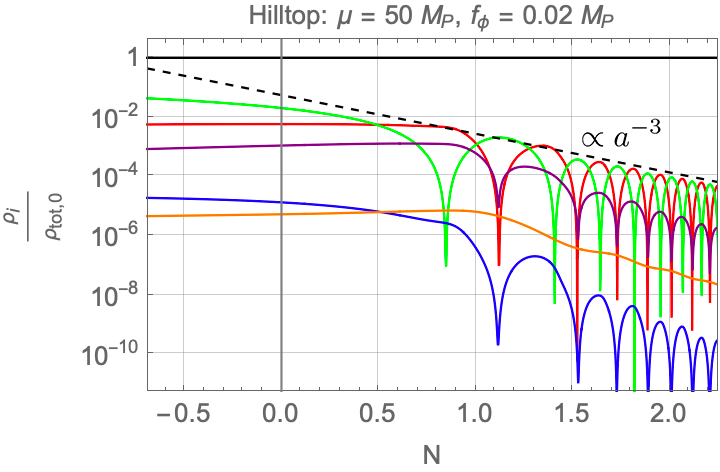} \hspace{5mm}
\includegraphics[width = 7cm]{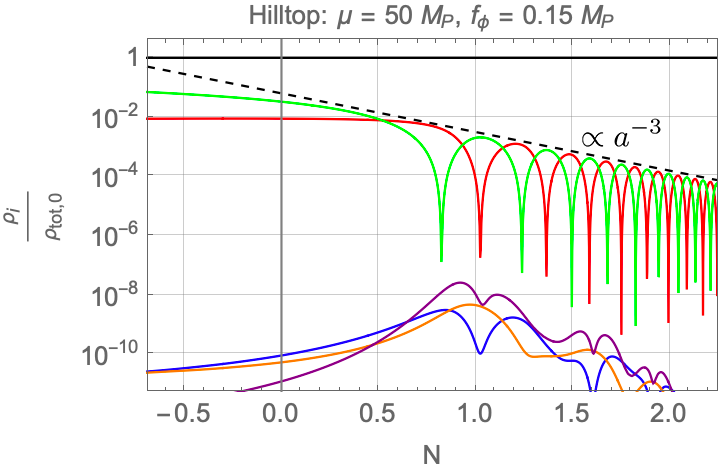} 
 \caption{\it Inflaton kinetic and potential energy density, as well as electric, magnetic and fermion energy density ratios to the initial total energy density of the Universe for the hilltop models with $\mu=10\,\Mp$ (upper panels) and $\mu=50\,\Mp$ (lower panels). The vertical gray lines display the value $a$ for which $\epsilon(a)=1$ and the dashed line shows the expected scaling of the dominant sector.
 \label{fig:Preheating-energy-breakdown-vB}}
\end{center}
\end{figure}

\subsection{Numerical results beyond the end of inflation}
Now that we have established a method to numerically compute the quantities
 $\rho_E$, $\rho_B$, $\rho_\psi$, $\mathcal H$ and $\mathcal G$, we aim to study the system evolution past $\epsilon=1$, and the onset of reheating. Indeed, the system (\ref{Staro-EoMs}) describes the most general interaction of the zero mode of both hypercharge gauge and inflaton fields. In particular, no assumption was made on the Universe geometry, hence there is no specific reason to stop its numerical computation at the end of inflation.
We will also find it convenient to present some numerical results using as the variable the number of $e$-folds before the end of inflation $N$, instead of the scale factor $a$, and related to it by
\be N = -\log{\frac{a_E}{a}} \ee
such that $N=0$ corresponds to the time $a_E$ when $\epsilon(a_E)=1$.

\begin{figure}[h]
\begin{center}
\includegraphics[width = 8.5cm]{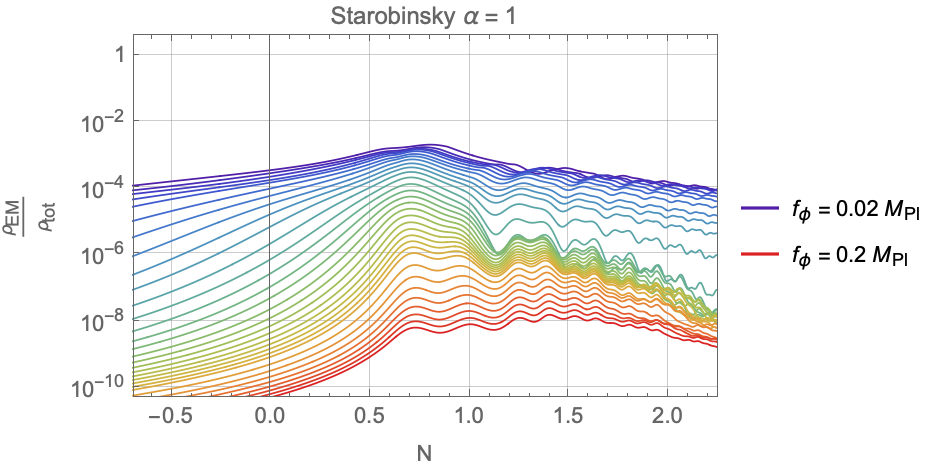}
\includegraphics[width = 6.5cm]{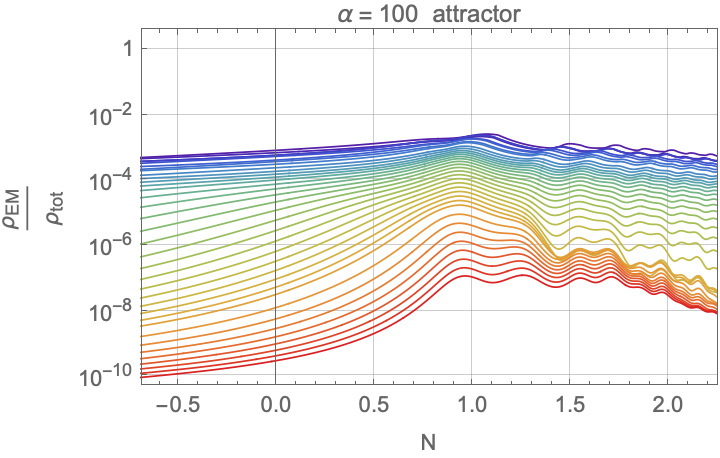} \\\vspace{2mm}
\includegraphics[width = 7.1cm]{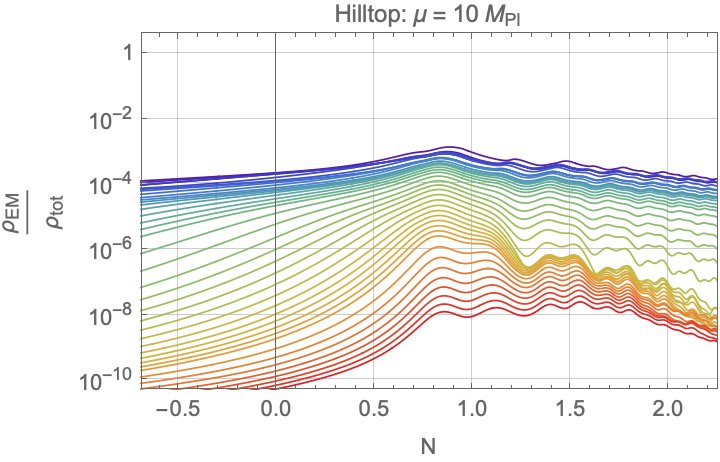}  \hspace{8mm}
\includegraphics[width = 7.1cm]{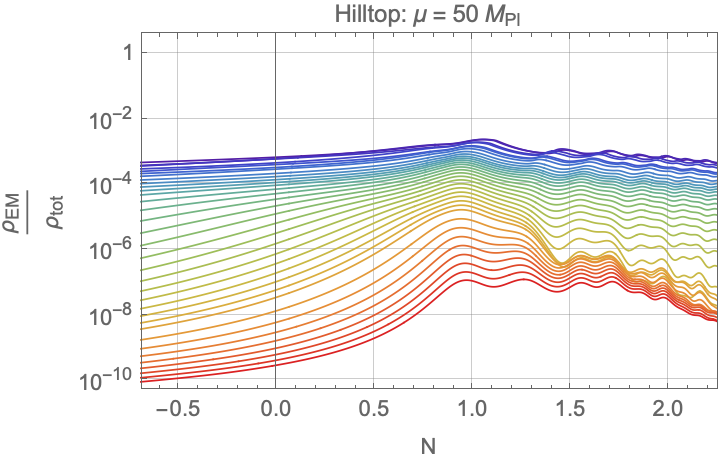} 
 \caption{\it Time evolution of the electromagnetic to total energy density fraction, during and after inflation for various values of the coupling $f_\phi$. The upper panels correspond to the $\alpha$-attractor model with $\alpha=1$ (top left) and $\alpha=100$ (top right) and the lower panels to the hilltop model with $\mu=10\,\Mp$ (bottom left) and $\mu=50\,\Mp$ (bottom right).
 \label{fig:Preheating-efficiency-1}}
\end{center}
\end{figure}

We show the postinflationary energy breakdown for selected value of $f_\phi$, for the $\alpha$-attractor models in Fig.~\ref{fig:Preheating-energy-breakdown-vA}, $\alpha=1$ (upper panels) and $\alpha=100$ (lower panels), and the hilltop models of inflation in Fig.~\ref{fig:Preheating-energy-breakdown-vB}, with $\mu=10\,\Mp$ (upper panels) and $\mu=50\,\Mp$ (lower panels). 
From the inflaton behavior, we see that the Universe enters a matter domination era as $\rho_\phi \sim a^{-3}$.
For high enough values of $f_\phi$, i.e.~$f_\phi \gtrsim 0.1\,\Mp$, we reproduce the results shown in Ref.~\cite{Cuissa:2018oiw}, whereas for $f_\phi \lesssim 0.1\,\Mp$ the electric and magnetic fields exhibit a different behavior: the former decays faster than the latter while oscillating. This is due to the fact, already mentioned in Ref.~\cite{Gorbar:2021rlt}, that the energy density for the electric component $\bm{E}=-\bm A'$ is much more sensitive to the Schwinger effect than the magnetic component $\bm{B}$, because it directly couples to the conductivity in the gauge field equation of motion (\ref{eq:Apm}).
On the other hand, the magnetic component reflects spatial effects, as it is defined by $\bm{B}=\bm{\nabla}\wedge\bm{A}$. In this work, we do not consider the inflaton spatial effects, $\bm{\nabla} \phi$, because this would require one to implement real fermion interactions in a lattice simulation. Hence, for low values of $f_\phi$, when the Schwinger effect is strongly affecting the system, the behavior of $\rho_B$ is expected to be subject to changes when the spatial effects are enabled; namely we expect to see a faster decay, like that of $\rho_E$. As also observed in Ref.~\cite{Gorbar:2021rlt}, the electric field, which is dominant during inflation, becomes subdominant afterwards.
Finally, we can see that for low values of $f_\phi$ the fermion energy density dominates the radiation energy density at the end of inflation as already highlighted in Ref.~\cite{Gorbar:2021rlt}.

The authors of Ref.~\cite{Cuissa:2018oiw} quote a \textit{sufficient criterion} for gauge preheating to happen, namely that at least an 80\% fraction of the total energy density of the Universe is electromagnetic energy. In the absence of the Schwinger effect, they found that this criterion is satisfied for values $f_\phi\lesssim 0.1\,\Mp$. However, as expected, the Schwinger effect significantly reduces the share of electromagnetic energy, as shown on Fig.~\ref{fig:Preheating-efficiency-1} for the considered models, which displays the ratio $\rho_{\rm EM}/\rho_{\rm total}$ for the four previous considered cases. We can see that the maximum is attained with a value $\sim 10^{-3}$, which precludes any gauge preheating, at least for $f_\phi \gtrsim 0.01\,\Mp$.
Another conclusion from Ref.~\cite{Cuissa:2018oiw} is that the spatial effects of the inflaton become relevant for sufficiently low values of $f_\phi$ and contribute to preheating. Since we are neglecting them in our simplified calculation, any negative statement concerning the possibility of gauge preheating due to the lack of enough electromagnetic energy should be a conservative one.
\begin{figure}[htb]
\begin{center}
\includegraphics[width = 10cm]{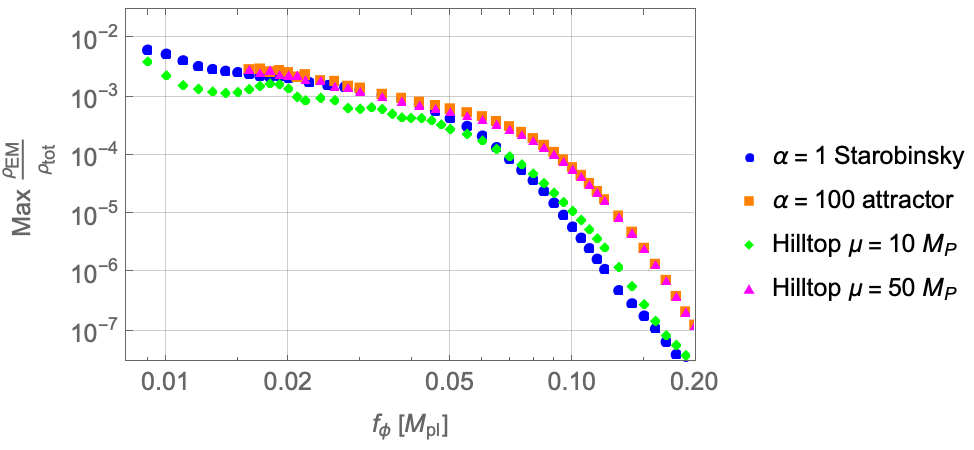} 
 \caption{\it Maximum value of the electromagnetic to total energy fraction as a function of $f_\phi$ for the four considered models: $\alpha$-attractor models, with  $\alpha = 1,\,100$, and hilltop models, $\mu=10,\,50\, \Mp$. Preheating seems unlikely to occur.
 \label{fig:preheating-potential_comp}} 
\end{center}
\end{figure}

The final results from our analysis can be summarized in Fig.~\ref{fig:preheating-potential_comp}, where we plot
the maximum value of the electromagnetic to total energy fraction as a function of $f_\phi$ (preheating efficiency) for the Starobinsky model, the $\alpha$-attractor model with $\alpha=100$ and the hilltop models with $\mu /\Mp=10,50$. For $f_\phi\gtrsim 0.01\,\Mp$, we obtain
\be
\frac{\rho_{\rm EM}}{\rho_{\rm tot}}\lesssim 0.01,
\ee
which seems to prevent gauge preheating as its efficiency is far from the value of $\sim 0.8$ established in the numerical analysis of Ref.~\cite{Cuissa:2018oiw}.

\subsection{End of reheating}
If gauge preheating does not occur, the inflaton will eventually decay by perturbative processes which depend on the inflaton total decay width $\Gamma_\phi$. Therefore at the time $t_{\rm rh}\sim 1/\Gamma_\phi$, the inflaton has completely decayed and the radiation domination era starts. 

Results from last sections have shown that shortly after inflation ends, the Universe is dominated by matter, hence we can approximate the Hubble parameter by
\be H \simeq \left(\frac{a_E}{a}\right)^\frac{3}{2} H_E, \hspace{1cm} H \simeq \frac{2}{3t},  \ee
where $H_E\equiv H(a_E)$, such that 
\be a_{\rm rh}\simeq a_E \left(\dfrac{3H_E}{2\Gamma_\phi} \right)^\frac{2}{3} \ee
is the end value after reheating by inflaton perturbative decays. Of course $a_{\rm rh}$ is a model-dependent quantity, which depends on the value of $\Gamma_\phi$, which in turn, depends on the couplings of the inflaton to the matter.

In particular, the coupling $1/f_\phi$ of the inflaton to the hypercharge Chern-Simons density provides a channel for the perturbative decay of the inflaton into a pair of hyperphotons $A$, as $\phi\to AA$. This decay has a width given by~\cite{Adshead:2015pva}
\be
\Gamma(\phi\to AA)\simeq\frac{m^3_\phi}{64\pi f^2_\phi}.
\ee
where $m_\phi$ is the inflaton mass given by 
\be  m^2_\phi = \left.\dfrac{\partial^2 V}{\partial \phi^2}\right|_{\phi=\phi_{\rm min}}.  \ee
For the $\alpha$-attractor (hilltop quartic) model, we have $\phi_{\rm min,\alpha}=0$ ($\phi_{{\rm min},\,h}=\mu$) and
\be m^2_{\phi,\, \alpha} = \dfrac{4 \Lambda_\alpha^4}{3 \alpha \Mp^2}, \hspace{2cm} m^2_{\phi,\, h} =  \dfrac{32 \Lambda_h^4}{\mu^2} .\ee
\indent
In the simplest case where the inflaton is only coupled to the hypercharge gauge bosons through the Chern-Simons density, the total width is $\Gamma_\phi=\Gamma(\phi\to AA)$.
Using the masses found above we have that 
\be\Gamma_\phi\simeq 12\ (3.0) \cdot 10^{-18}\cdot \dfrac{\Mp^3}{f_\phi^2},
\label{eq:Gamma-alpha}
\ee
 for $\alpha = 1 \ (100)$ in the $\alpha$-attractor models, and
\be
 \Gamma_\phi\simeq 4.2\ (21) \cdot 10^{-19}\cdot \dfrac{\Mp^3}{f_\phi^2},
 \label{eq:Gamma-hilltop}
  \ee
for $\mu = 10 \ (50)\,\Mp$ in the hilltop models. 
%
%
\newline\indent
The value of the scale factor and the temperature at reheating, $a_{\rm rh}$ and $T_{\rm rh}$, are given by
\be
\frac{a_{\rm rh}}{a_E}\simeq 0.4 \left( \frac{T_{\rm rh}}{T_{\rm rh}^{\rm ins}}\right)^{-4/3}, \hspace{2cm}
 \frac{T_{\rm rh}}{T_{\rm rh}^{\rm ins}}\simeq \sqrt{\frac{\Gamma_\phi}{H_E}}.
\ee
Consequently we can express $a_{\rm rh}$ and $T_{\rm rh}$ as functions of all the involved parameters, namely $f_\phi$, and $\alpha$ ($\mu$) for $\alpha$-attractor (hilltop quartic) model.
%
In particular, the relevant parameter for baryogenesis is the ratio $T_{\rm rh}/T_{\rm rh}^{\rm ins}$ given by
\be
\frac{T_{\rm rh}}{T_{\rm rh}^{\rm ins}}  \simeq 1.9\ (0.8) \cdot 10^{-4}\, \left(\dfrac{0.01}{f_\phi/\Mp}\right),   
\label{eq:TRHalpha}
\ee
 for $\alpha = 1\ (100)$ in the $\alpha$-attractor models, and
\be
\frac{T_{\rm rh}}{T_{\rm rh}^{\rm ins}}  \simeq 0.4\ (0.7) \cdot 10^{-4}\left(\dfrac{0.01}{f_\phi/\Mp}\right) ,
\label{eq:TRHhilltop}
\ee
for $\mu = 10 \ (50)\,\Mp$ in the hilltop models. As we will see in the next section the obtained values of the ratio $T_{\rm rh}/T_{\rm rh}^{\rm ins}$ are fully consistent with the general baryogenesis results, see Fig.~\ref{fig:baryogenesis}, provided that $f_\phi \lesssim 0.03\, \Mp$.
\newline\indent
In the presence of extra couplings of the inflaton to matter, the predictions for the inflaton decay width, Eqs.~(\ref{eq:Gamma-alpha}) and (\ref{eq:Gamma-hilltop}), and the reheating temperature,  Eqs.~(\ref{eq:TRHalpha}) and (\ref{eq:TRHhilltop}), will change in a model-dependent way, as well as the model predictions concerning the generation of the baryon asymmetry.
\newline\indent
Of course in the hypothetical case where the explosive production of gauge fields should have prevailed over the perturbative inflaton decays, gauge preheating would have taken place over a few $e$-folds after the end of inflation.
 As we see from the previous results, this is never the case and gauge preheating is never strong enough to reheat the Universe after the period of cosmological inflation. This result does not preclude that, in the presence of a strong coupling $\lambda$ of the inflaton with some other field, e.g.~a scalar (or a fermion), there could exist an explosive production of that scalar (or fermion), triggering preheating of the Universe after inflation~\cite{Cosme:2022htl}.

\subsection{Baryon asymmetry}
\label{sec:bau}

Before concluding this paper we wish to make a small comment on the baryogenesis issue at the electroweak phase transition. In Ref.~\cite{Cado:2022evn}, we presented a model of inflation that leads to a successful BAU. The effective potential for the inflaton, labeled therein as $\chi$, was the Starobinsky potential\footnote{In fact, we used in Ref.~\cite{Cado:2022evn} an scalar field $\phi$ non-minimally coupled with gravity as $\mathcal L=-\frac{1}{2}g\phi^2 R+\dots$, which yields for the canonically normalized field $\chi$ in the Einstein frame an $\alpha$-attractor potential with $\alpha = 1+ \tfrac{1}{6g} \in [4.3,17.6]$, where the lower bound was coming from imposing the naive unitarity bound $g\phi^2<\Mp^2$. As the dependence in $\alpha$ (hence in $g$) is tiny, we choose to show in the present paper the result for $\alpha=1$, hence for the Starobinsky potential (which would correspond in Ref.~\cite{Cado:2022evn} to the limit $g\gg 1$).}, and we did consider the Schwinger equilibrium and maximal estimates. Hence it is straightforward, using our numerical analysis in this paper, to make an update of the final results for the BAU for inflation driven by the $\alpha$-attractor models with $\alpha=1$.

As all details are explained in Secs.~6 and~7 of Ref.~\cite{Cado:2022evn}, we skip them here and go straight to the final result. First of all we show in Fig.~\ref{fig:baryogenesis} the analogous plot to Fig.~9 of Ref.~\cite{Cado:2022evn}, namely the parameter space that provides a successful BAU. 
\begin{figure}[htb]
\begin{center}
\includegraphics[width = 11cm]{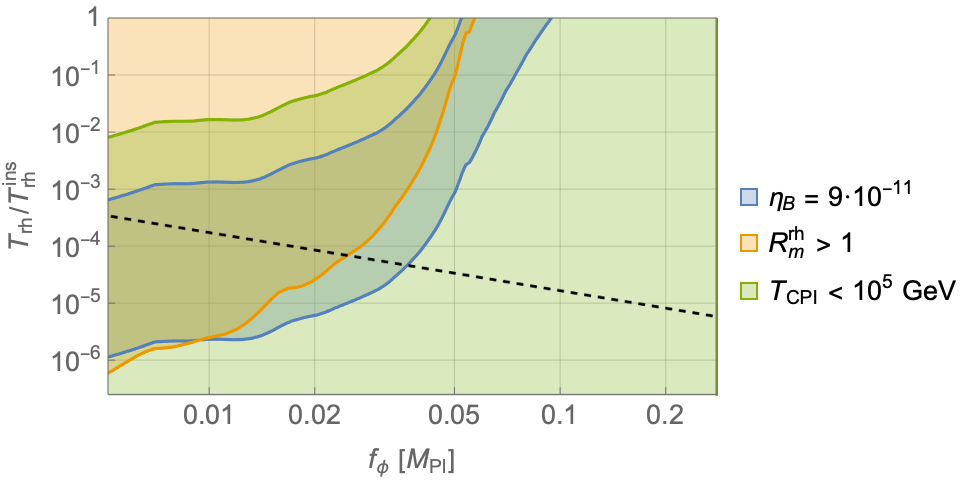} 
 \caption{\it The baryogenesis window in the  parameter space $(f_\phi,T_{\rm rh}/T_{\rm rh}^{\rm ins})$ for the Starobinsky potential ($\alpha$-attractor model with $\alpha=1$). The dashed line corresponds to Eq.~(\ref{eq:TRHalpha}).}
  \label{fig:baryogenesis}
\end{center}
\end{figure}
In particular, we display in blue the region where the asymmetry parameter meets its observational value given by
\be\eta_B  \simeq   4 \cdot 10^{-12} \, f_{\theta_W}  \frac{\mathcal{H}}{H_E^3} \left( \frac{H_E}{10^{13} \, \text{GeV}} \right)^{\frac{3}{2}} \left(\frac{T_{\rm rh}}{T_{\rm rh}^{\rm ins}}  \right) \, \, \simeq \, 9 \cdot 10^{-11} ,\label{constraint-nB} \ee
where we have imposed the observed value~\cite{Zyla:2020zbs} in the right-hand side. Following Refs.~\cite{Domcke:2019mnd, Cado:2021bia} we define the parameter $f_{\theta_W}$, which encodes all the details of the EW phase transition and its uncertainties, as 
\be f_{\theta_W}  = -\sin (2 \theta_W) \, \dfrac{d\theta_W}{d\ln T}\bigg\rvert_{T=135\text{ GeV}}, \quad 5.6 \cdot 10^{-4} \lesssim f_{\theta_W}  \lesssim 0.32. 
\label{eq:ftheta}
\ee

In addition to their dependence on the gauge sector observables, the quantities used in this section vary according to the ratio of the reheating over the instant reheating temperature. This parameter hence adds to $f_\phi$ in the parameter space. The reheating temperature is computed as
\be
T_{\rm rh}= \left(\frac{90}{\pi^2 g_\ast}\right)^{\frac{1}{4}} \sqrt{\Gamma_\phi \Mp}, 
\label{eq:Trh}
\ee
where $g_\ast=106.75$ is the SM number of relativistic degrees of freedom, and we define $T_{\rm rh}^{\rm ins}$ as a reference temperature given by the above equation with $\Gamma_\phi\simeq H_E$, which is obtained from the simulation. It would correspond to the reheating temperature for instant reheating, and takes the value $T_{\rm rh}^{\rm ins}\simeq 2.87 \cdot 10^{15}$~GeV.
%
%
%
%
Using Eq.~(\ref{eq:TRHalpha}) it is possible to link the reheat temperature to the parameter $f_\phi$. The corresponding plot is shown in Fig.~\ref{fig:baryogenesis} which shows that it provides a wide window for baryogenesis.

Second, we display in orange the region where the magnetic Reynold's number at reheating $\mathcal R_m^{\rm rh}$ is bigger than one, hence ensuring that the required magnetohydrodynamical conditions are fulfilled for the (hyper)magnetic fields to survive until the electroweak crossover. As we are in the viscous regime, it can be computed as~\cite{Cado:2022evn}
\be\mathcal{R}_m^{\rm rh}  \approx 5.9 \cdot 10^{-6} \; \frac{\rho_{B} \ell_{B}^2}{H_E^2}  \left( \frac{H_E}{10^{13} \, \text{GeV}} \right) \left(\frac{T_{\rm rh}}{T_{\rm rh}^{\rm ins}}  \right)^{\frac{2}{3}},\label{constraint-Rm}\ee
where $\ell_{B}$ is the physical correlation length of the magnetic field given by
\be \ell_{B}  =\frac{2\pi}{\rho_B\,a^3}  \int^{k_c}d k \, \frac{k^3 }{4\pi^2} \left(|A_+|^2+|A_-|^2\right),\ee
which can be numerically computed during the simulation in the same way as the other observables.

Third, and last\footnote{Besides, we checked that the generation of baryon isocurvature perturbation provides no constraint.}, we show in green the condition on the chiral plasma instability (CPI) temperature, ensuring that the CPI time scale is long enough to allow all right-handed fermionic states to come into chemical equilibrium with the left-handed ones via Yukawa coupling interactions (so that sphalerons can erase their corresponding asymmetries in particle number densities) before CPI can happen. The estimated temperature at which CPI takes place is
\be T_{\rm CPI}/\textrm{GeV}  \approx    4 \cdot 10^{-7} \,  \; \frac{\mathcal{H}^2}{H_E^6} \, \left( \frac{H_E}{10^{13} \, \text{GeV}} \right)^3\left(\frac{T_{\rm rh}}{T_{\rm rh}^{\rm ins}}  \right)^2 \,.
\label{constraint-TCPI}\ee
The constraint $T_{\rm CPI}\lesssim 10^5$~GeV (the temperature at which $e_R$ comes into chemical equilibrium) guarantees that the CPI cannot occur before the smallest Yukawa coupling reaches equilibrium and all particle number density asymmetries are erased, preventing thus the cancellation of the helicity generated at the reheating temperature.

Therefore, as we can see from Fig.~\ref{fig:baryogenesis}, the resulting baryogenesis window for the Starobinsky potential is close to the Schwinger equilibrium estimate for $f_\phi \lesssim 0.06\,\Mp$, just as the corresponding results on helicity and magnetic energy density suggest (see the green dots of Fig.~\ref{fig:inflaton-result-fphi}). However, for $f_\phi \gtrsim 0.06\,\Mp$ there is no space for the BAU, as the production of gauge fields is too weak, unlike in the previous results from Ref.~\cite{Cado:2022evn}. 
In addition to this we have seen that the reheating temperature is constrained by the model, see Eq.~(\ref{eq:TRHalpha}), as we can see from Fig.~\ref{fig:baryogenesis} and compatibility of the model reheating temperature with the baryogenesis results translates into the baryogenesis region on the parameter $f_\phi$ 
\be
f_\phi\lesssim 0.03\, .\Mp
\ee 
%
Finally, one of the results of this paper is then that baryogenesis at the electroweak phase transition is favored by low reheating temperatures, in the range $10^{-6}\,T_{\rm rh}^{\rm ins}\lesssim T_{\rm rh}\lesssim 10^{-3}\,T_{\rm rh}^{\rm ins}$.

\section{Conclusions}
\label{sec:conclusions}

In this paper, we have studied by means of numerical computations the effect of the Schwinger particle production on the helical hypermagnetic fields produced at the end of inflation. 
The inflaton field $\phi$ can decay, through its coupling to the Chern-Simons density $\tfrac{\phi}{4f_\phi}Y_{\mu\nu}\tilde{Y}^{\mu\nu}$, into helical hypermagnetic fields in a nonperturbative process.
When exiting the vacuum, the gauge modes are strong enough to create particle/antiparticle pairs of light fermions, which contribute to the electrical conductivity of the plasma.
The backreaction of fermion currents on the produced gauge fields acts as a damping force in the explosive production of helical gauge fields. This effect, called \textit{Schwinger effect}, was already considered in numerous studies of inflation and/or baryogenesis, where some analytical and numerical estimates were computed, mainly in configuration space while our calculation is done in momentum space.

The equations of motion are in fact a nontrivial integro-differential system. It was solved numerically by using a fourth order Runge-Kutta method, with details being displayed in the Appendices. The computed observables of interest are the electric and magnetic energy density, the helicity as well as the helicity time derivative. We assumed a homogeneous inflaton with only zero mode, hence we did not treat any spatial effects. Besides, we also ensured the convergence of the algorithm and its invariance to the initial conditions. 

First of all we have checked that we recover previous results in the slow roll inflation regime by making the same approximations required by an analytical resolution. In this way, we validate our code, i.e.~we verify that our code produces the right results in known cases such as the backreactionless case, where the Schwinger effect is turned off, and the gradient expansion formalism, where the Bunch-Davies parameter $\Delta$ was first introduced.

In a second step, still in the slow roll regime, we considered a specific model of inflation, namely the Starobinsky potential, in order to account for the instability parameter as a function, $\xi(a)$, instead of the constant imposed by the analytical approximations. That way, we could also implement the effects of a function $\Delta(a)$ obtained from the plasma evolution on the gauge production itself. 

We then simulated, in a third step, the full system, where neither the slow roll conditions nor the Universe geometry (e.g.~de Sitter) are imposed. In order words, the inflaton equation of motion was computed alongside with the gauge one, taking the backreaction of the latter to the former into account along with the Schwinger effect. We compare our result to the previous setup and found perfect agreement as long as the slow roll conditions are met. When inflation is near its end, the full solution diverges from the slow roll results and produces, as expected, less energy density and helicity. 

Finally we will comment on the implication about two related topics: gauge preheating and baryogenesis.
As our code is free from any geometrical issues, and only requires a model of inflation, we let the simulations run until the onset of reheating to compute the electromagnetic to total energy density ratio. We choose two well-known classes of models that satisfy the cosmological constraints as illustrative examples. 
Previous studies have quoted a sufficient criterion for gauge preheating to happen, namely that this fraction should be at least $\gtrsim$ 80\%~\cite{Cuissa:2018oiw}. However, our numerical estimates suggest that the Schwinger effect significantly reduces the share of electromagnetic energy for the considered models and preheating is unlikely to occur.
Moreover, since we are neglecting all spatial effects, any negative statement concerning the possibility of gauge preheating due to the lack of electromagnetic energy should be a conservative one. On the other hand our results do apply to the considered class of inflationary models. They show a certain degree of model dependence, so we cannot exclude a qualitatively different result for models of inflation other than the considered ones.

On the other hand, as a successful baryogenesis does depend on a delicate equilibrium between the amount of helicity, magnetic energy density, and magnetic correlation length, damped fields do not necessarily mean no baryon asymmetry in the late Universe. Actually, as a result of our numerical calculation, we have found there is still a window in the parameter space for baryogenesis to happen as long as $f_\phi \lesssim 0.05\,\Mp$, while consistency from the perturbative decay channel of the inflaton into hypergauge bosons implies the bound $f_\phi\lesssim 0.03\,\Mp$. Moreover, baryogenesis is favored for low enough values of the reheating temperature $T_{\rm rh}\lesssim 10^{-3}\, T_{\rm rh}^{\rm ins}$. Of course, the baryogenesis predictions should, to some extent, depend on the model of inflation. In this way, our result here is restricted to the Starobinsky model and should be considered just as a ``proof of existence" for baryogenesis in the presence of the Schwinger effect.

These two comments should be viewed as hints for future studies that address the production of gauge fields at the end of inflation. 
Of course, a full lattice simulation of the Schwinger effect involving fermions remains to be done.

\vspace{0.5cm}
\section*{Acknowledgments}
YC would like to thank Adrien Florio, for interesting discussions at the early stage of this work. MQ is indebted to Daniel Figueroa for interesting discussions, full of new ideas, on the subject. This work is supported by the Departament d'Empresa i Coneixement, Generalitat de Catalunya Grant No.~2017SGR1069, by the Ministerio de Economía y Competitividad Grant No.~FPA2017-88915-P. IFAE is partially funded by Centres de Recerca de Catalunya. YC is supported by the European Union’s Horizon 2020 research and innovation programme under the Marie Sklodowska-Curie Actions No.~754558.

\appendix

\section{Numerical method: the slow roll case}
\label{app:slow-roll}
We provide here the technical details for the solution of Eq.~(\ref{eq:Apm-vs-a}) subject to the initial condition (\ref{BD-modes-a}). For convenience, we implement the numerical computation in units of $H_E$. Writing
\be x_\lambda(a) =  A_\lambda (a), \hspace{2cm} y_\lambda(a) =\frac{d A_\lambda}{d a}(a), \ee
Eq.~(\ref{eq:Apm-vs-a}) becomes the following system:
\be \frac{d}{da} \left(\begin{array}{c}x_\lambda \\y_\lambda\end{array}\right) = \left(\begin{array}{cc}0 & 1 \\\frac{k}{a^3}\left(2\lambda \xi -\frac{k}{a} \right) & -\frac{1}{a}\left(\frac{\sigma}{a}+2\right)\end{array}\right) \left(\begin{array}{c}x_\lambda \\y_\lambda\end{array}\right) \quad \Leftrightarrow \quad \frac{d \bm{x}}{da}= \bm{f}(a, \bm{x}).\ee

To perform each time step $ \Delta a$, we use the fourth order Runge-Kutta (\texttt{RK4}) algorithm:
\begin{subequations} \begin{eqnarray}
\bm{\lambda}_1 &=& \bm{f} (a_i, \bm{x}_i) \\
\bm{\lambda}_2 &=& \bm{f} \left(a_i +\tfrac{1}{2} \Delta a, \bm{x}_i+\tfrac{1}{2}\Delta a\bm{\lambda}_1\right) \\
\bm{\lambda}_3 &=&  \bm{f} \left(a_i +\tfrac{1}{2} \Delta a,  \bm{x}_i+\tfrac{1}{2}\Delta a\bm{\lambda}_2\right) \\
\bm{\lambda}_4 &=& \bm{f} \left(a_i +\Delta a, \bm{x}_i+\Delta a\bm{\lambda}_3\right) \\
a_{i+1}&=&a_i + \Delta a \\
\bm{x}_{i+1} &=&\bm{x}_{i}+ \tfrac{1}{6} \Delta a(\bm{\lambda}_1+2\bm{\lambda}_2+2\bm{\lambda}_3+\bm{\lambda}_4) \end{eqnarray} \label{eq:RK4-aglorithm} \end{subequations}
Note that $\bm{x}$ is complex, hence we solve the above system for both real and imaginary parts but with their specific initial conditions. These are mode-dependent as it takes longer for modes with bigger wave number to leave the BD vacuum. Therefore, we choose as initial condition for each mode
\be a_{k,0} =\frac{k}{x_{\rm BD}}, \ee
where we choose the factor $x_{\rm BD}$ in order to make sure that we initialize the gauge field sufficiently deep inside the Hubble radius. Its exact value is subject to analysis and is discussed later. As we can see from (\ref{def:energy-densities}) and (\ref{def:AB-and-EB}), high values of $k$ are dominating the integral hence large modes are negligible compared to small ones. This makes us to choose a lower bound on the $k$ range such that the initial time of the simulation is 
\be a_0 =\frac{k_{\rm min}}{x_{\rm BD}} . \label{BD-penetration}\ee
In that way, at $a_0$ we make sure that all the modes are in their respective vacua, which implies $\sigma=0$ as explained above. 

In practice, this means that the modes with $k>x_{\rm BD}a$ are given by the following relations
\bse 
 {\rm Re} (x_{\lambda,i}^{\rm BD}) &=& \sqrt{\frac{\Delta_i}{2k}} \cos \frac{k}{a_i}, \\
  {\rm Im} (x_{\lambda,i}^{\rm BD}) &=& \sqrt{\frac{\Delta_i}{2k}} \sin \frac{k}{a_i}, \\
 {\rm Re} (y_{\lambda,i}^{\rm BD}) &=& \frac{1}{a_i^2} \sqrt{\frac{\Delta_i}{2}} \left( \sqrt{k} \sin \frac{k}{a_i}-\frac{\sigma_i}{2\sqrt{k}} \cos \frac{k}{a_i} \right), \\
 {\rm Im}  (y_{\lambda,i}^{\rm BD})&=& \frac{1}{a_i^2} \sqrt{\frac{\Delta_i}{2}}  \left(- \sqrt{k} \cos \frac{k}{a_i}-\frac{\sigma_i}{2\sqrt{k}} \sin \frac{k}{a_i} \right)\ese
 while the others are evolving with the \texttt{RK4} algorithm.

The time steps are distributed on a logarithmic scale
\be \log{a_{i}} - \log{a_{i-1}}= \log{a_{i+1}} - \log{a_{i}} ,\ee
so that the discretization is the same for each order of magnitude. This means $\Delta a$ grows exponentially with $a$. The advantage of this method is that there is a refinement of the grid for small values of $a$, at the beginning of inflation. The same is done for the discretization in $k$.

We explored the numerical convergence of the solution, both in the number of $a_i$'s, labeled as $N_a$, and in the number of $k_j$'s, labeled as $N_k$.
Provided that $N_a >2000$ and $N_k>200$, the simulations are very stable and the output does not depend on the discretization. For big values of $f_\phi$, $f_\phi \gtrsim 0.1$, we can even lower the number of time steps needed.

Besides, we must choose the BD penetration factor $x_{\rm BD}$ such that it produces trustable results.
We have done a numerical analysis and conclude that depending on the value of $N_a$, a range $20<x_{\rm BD}<50$ yields trustable results. We hence choose throughout this work the following values
\be x_{\rm BD}=20, \hspace{1cm} N_a=500,\,1000,\,2000,\hspace{1cm} N_k=300. \ee

At each time step, we compute the electric and magnetic energy density as
\begin{subequations} \begin{eqnarray}
\rho_{E}^{i} &=&  \int_{k_{\rm min}}^{k_c^i} d k \,  \frac{k^2}{4 \pi^2}\left(| y_{i}^+(k)|^2 + | y_{i}^-(k)|^2\right), \\
\rho_{B}^{i} &=& \frac{1}{a_i^4} \int_{k_{\rm min}}^{k_c^i} d k \,  \frac{k^4}{4 \pi^2}\left(|x_{i}^+(k)|^2 + |x_{i}^-(k)|^2\right),\end{eqnarray} \label{rho-discretized} \end{subequations}
where we choose 
\be k_c^i = a_i\xi+ \sqrt{  \left(a_i\xi\right)^2+ \frac{a_i^2}{2}  \left[\frac{\sigma_i-\sigma_{i-1}}{a_i-a_{i-1}}+\frac{\sigma_i}{a_i}\left(\frac{\sigma_i}{2a_i}+1\right) \right]}. \label{cutoff-num}\ee
such that we cut off the spectra to retain only modes outside the horizon. The helicity (\ref{ABdef}) and its derivative (\ref{ABtoEB}) become
\bse \mathcal{H}_{i}  &=&  \frac{1}{a_i^3}  \int_{k_{\rm min}}^{k_c^i} d k \,  \frac{k^3}{2 \pi^2}\left(|x_{i}^+(k)|^2 - |x_{i}^-(k)|^2\right), \\
\mathcal{G}_{i}  &=&  \frac{1}{a_i^2} \int_{k_{\rm min}}^{k_c^i} d k \,  \frac{k^3}{2 \pi^2}\left(|x_{i}^+(k)y_{i}^+(k)| - |x_{i}^-(k)y_{i}^-(k)|\right). \end{eqnarray} \label{hel-discretized} \end{subequations}
In the numerics, these integrals are performed numerically over the range of $k$ that takes $N_k$ discrete values. 
If the Schwinger effect is taken into account, we turn on the possibility of having $\sigma$ computed at each time step $a_i$ of the numerical computation with
\be\sigma_{i+1} = \frac{41 \,g'^3}{72 \,\pi^2}\, a_i\,\sqrt{2\rho_{B}^i}\;\coth\left(\pi\sqrt{\frac{ \rho_{B}^i}{\rho_{E}^i}} \right) \label{sigma-num-not0} \ee
and injected into the calculation of the next step. Otherwise, we keep it zero.
Last, the fermion energy density is computed as
\be \rho_{\psi}^{i} = \frac{\sigma_i}{a_i^2}  \int_{k_{\rm min}}^{k_c^i} d k \,  \frac{k^2}{\pi^2}\sum_{\lambda=\pm} \left[{\rm Re}(x_i^\lambda){\rm Re}(y_i^\lambda)+{\rm Im}(x_i^\lambda){\rm Im}(y_i^\lambda)\right]. \ee
Finally, we stop the simulation at $a=a_E$. Quantities at that time are compared to the known analytical results. The color matching dashed vertical lines in Fig.~\ref{fig:spectra-many-a} show the cutoff values $k_c^i$ computed from (\ref{cutoff-num}). They agree perfectly with the point where the BD vacuum modes become dominant for large $k$. 

\section{Numerical method: full analysis}
\label{app:full}
The numerical implementation follows from the previous case. Defining the variables
\be w=\phi, \qquad x=\frac{d\phi}{d a}  , \qquad y_\lambda = A_\lambda, \qquad z_\lambda =\frac{d A_\lambda}{d a} \ee
we transform the above coupled system of differential equations (\ref{Staro-EoMs}) into the system
\bse \frac{d w}{da}  &=& x  \\
\frac{d x}{da}  &=& -\dfrac{\mathcal{G}}{a^2 H^2 f_\phi}-\frac{4-\mathcal{F}}{a} \;x-\frac{V'(w)}{a^2H^2 }\\
\frac{d y_\lambda}{da}  &=& z_\lambda \\
\frac{dz_\lambda }{da}  &=& \frac{k}{a^2 H}\left(\frac{\lambda}{f_\phi}\; x -\frac{k}{a^2 H}\right) \,y_\lambda - \frac{1}{a}\left(2 -\mathcal{F}+\frac{\sigma}{aH}\right)\; z_\lambda \ese
which is equivalent to writing
\be \frac{d \bm{x}}{da}= \bm{f}(a, \bm{x}). \ee
We recall that $w,\, x \in \mathbb{R}$ and $y_\lambda,\, z_\lambda \in \mathbb{C}$.
Similarly to the previous calculation with the slow roll approximation, we use the \texttt{RK4} algorithm (\ref{eq:RK4-aglorithm}) with the values of $H$, $\sigma$, $\mathcal{F}$ and $\mathcal{G}$ computed at each time step.

Inflaton initial condition could be set to
\be w_0 = \phi_\ast, \hspace{2cm} x_0=0.   \ee
However, the number of $e$-folds sets the initial time as $a_0={\rm e}^{-|N_\ast|} \sim 10^{-26}$, which is too small a number for the numerical implementation.
We then proceed as follows. For $a \lesssim  k_{\rm min}/x_{\rm BD}$, and sufficiently low $k_{\rm min}$, the gauge field modes stay in their vacuum and the total contribution to $\Delta(a)$ is negligible. Hence we do not need to perform the numerical simulation before that time, as the inflaton is the main player, so we can solve its equation of motion analytically. Instead, we fix the start of the simulation like before, at $a_0 = k_{\rm min}/x_{\rm BD}$ and we compute the corresponding number of $e$-folds $N$ which leads us to the corresponding value of $\phi(N)$. Therefore, the initial condition must be set to $w_0$ such that
\be  \int_{\phi_E}^{w_0} \dfrac{V(\phi)}{V'(\phi)} \,d \phi=-\Mp^2\log{a_0} \ee 
and, using $\dot{\phi}\simeq -\frac{V'(\phi)}{3H}$ which is valid at the early stages of inflation, 
\be x_0 = - \dfrac{V'(w_0)}{3a_0H_0^2}.\ee
As for the gauge field, initial conditions are set in the same way as in the slow roll approximation, see App.~\ref{app:slow-roll}.

\bibliographystyle{JHEP}
\bibliography{refs}

\end{document}